\RequirePackage{lineno} 
\documentclass[aps,twocolumn,prd,superscriptaddress,showkeys,floatfix]{revtex4-2}

\usepackage{graphicx,color}
\usepackage{dcolumn}
\usepackage{bm}
\usepackage{appendix}
\usepackage{listings}
\usepackage{bbold}
\usepackage{comment}
\usepackage{overpic}
\usepackage{amsmath}
\usepackage{amssymb}
\usepackage{rotating}
\usepackage[colorlinks,linkcolor=blue,anchorcolor=blue,citecolor=blue]{hyperref}

\def\be{\begin{equation}}
\def\ee{\end{equation}}
\def\M{\mathcal{M}}

\begin{document}

\voffset 1.0cm

\title{\Large K-matrix Analysis of $e^+e^-$ Annihilation in the Bottomonium Region}

\author{N. H\"{u}sken}
\affiliation{Department of Physics, Indiana University, Bloomington, Indiana 47405, USA.}
\affiliation{Johannes Gutenberg University of Mainz, Johann-Joachim-Becher-Weg 45, D-55099 Mainz, Germany.}

\author{R.E. Mitchell}
\affiliation{Department of Physics, Indiana University, Bloomington, Indiana 47405, USA.}

\author{E.S. Swanson}
\affiliation{Department of Physics and Astronomy, University of Pittsburgh, Pittsburgh PA 15260, USA. }

\begin{abstract}

We perform the first global and unitary analysis of $e^+e^-\to b\bar{b}$ cross sections.  We analyze exclusive cross sections in the $B\bar{B}$, $B^* \bar{B}(+c.c.)$, $B^*\bar{B}^*$, $B_s^*\bar{B}_s^*$, $\Upsilon(nS)\pi^+\pi^-$ and $h_b(nP)\pi^+\pi^-$ channels as well as the total inclusive cross section for $b\bar{b}$ production. Pole positions and residues are determined for four vector states, 
which we associate with 
the $\Upsilon(4S)$, $\Upsilon(10750)$, $\Upsilon(5S)$ (or $\Upsilon(10860)$), and $\Upsilon(6S)$ (or $\Upsilon(11020)$).  We find strong evidence for the new $\Upsilon(10750)$ recently claimed by Belle, although with parameters not well constrained by the data. Results presented here cast doubt on the validity of branching ratios reported earlier using Breit-Wigner parameterizations or ratios of cross sections.
We also compare our results with
a selection of theoretical calculations for the vector bottomonium spectrum.

\end{abstract}

\maketitle

\section{Introduction}

The spectrum of vector~($J^{PC} = 1^{--}$) bottomonium states above $B\bar{B}$ threshold
has been the source of a series of surprises and unresolved issues.
The initial exploration of this region using inclusive $e^+e^-$ annihilation to hadrons~\cite{BESSON85,LOVELOCK85} 
showed evidence for the 
production of two states with masses heavier than the $\Upsilon(4S)$, consistent with
potential model expectations for the $\Upsilon(5S)$ and $\Upsilon(6S)$. 
More recent measurements of the same process~\cite{AUBERT09E,SANTEL16,DONG20A}
have revealed more complex structure.  
While the putative $\Upsilon(5S)$ and $\Upsilon(6S)$ states 
(also called the $\Upsilon(10860)$ and $\Upsilon(11020)$, respectively)
still appear as prominent peaks in the inclusive cross section,
the effects due to coupled-channel scattering and the opening of a variety of open bottom thresholds
(e.g., $B^*\bar{B}$, $B^*\bar{B}^*$, $B_s \bar{B}_s$, $B_s^*\bar{B}_s$, $B_s^* \bar{B}_s^*$)
are now more apparent and complicate the observed spectrum.
Extracting vector bottomonium masses, total widths, and partial $e^+e^-$ widths from these spectra
has posed serious challenges.
While fits to the inclusive $e^+e^-$ spectrum using a coherent sum of Breit-Wigner amplitudes are
possible~\cite{DONG20A}, the fits violate unitarity and the results are expected to be unreliable. 

Recent measurements of the energy dependence of exclusive $e^+e^- \to B^{(*)}\bar{B}^{(*)}$ cross sections~\cite{MIZUK21bb}
confirm the  importance of coupled-channel scattering.  
Rather than showing distinct peaks for the $\Upsilon(5S)$ and $\Upsilon(6S)$, 
the cross sections are
marked by dramatic peaks and valleys at various open-bottom thresholds.
These non-trivial features in the open-bottom cross sections undermine older measurements of the $\Upsilon(5S)$ and 
$\Upsilon(6S)$ branching fractions,
such as those currently listed
in the Particle Data Group's Review of Particle Properties~(RPP)~\cite{pdg}.
Previous branching fractions of the $\Upsilon(5S)$ to open-bottom final states, for example,
were estimated by first measuring the cross section of $e^+e^-$ to a given open-bottom final state 
at an energy near the presumed mass of the $\Upsilon(5S)$ and 
then dividing by the inclusive $b\bar{b}$ cross section at the same energy~\cite{AQUINES06,HUANG07,DRUTSKOY10}.
This ratio of cross sections would approximate an $\Upsilon(5S)$ branching fraction 
only if the $\Upsilon(5S)$ were produced in isolation, an assumption we now know to be false.

Besides strong coupled-channel effects in the open-bottom final states, 
anomalously large cross sections for $e^+e^-$ to closed-bottom channels, such as $\pi\pi\Upsilon(nS)$ ($n=1,2,3$) and $\pi\pi h_b(nP)$ ($n=1,2$), have been observed~\cite{CHEN08,CHEN10,ADACHI12,MIZUK16,MIZUK19}.
Their production rates were later found to be enhanced by the presence of the exotic isovector bottomonium-like states, the $Z_b(10610)$ and $Z_b(10650)$~(also called the $Z_b$ and $Z_b^\prime$), which decay to $\pi\Upsilon(nS)$ and $\pi h_b(nP)$~\cite{BONDAR12,GARMASH15}.
In contrast to the complications in the open-bottom channels, the $\Upsilon(5S)$ and $\Upsilon(6S)$ appear to be well-isolated in the closed-bottom channels, which allows for a more reliable extraction of their mass and width~\cite{MIZUK19}.  
These relatively well-behaved cross sections also provide evidence for an additional state, the $\Upsilon(10750)$~\cite{MIZUK19}, which may be the $\Upsilon(3D)$ bottomonium state.

With the recent publication of inclusive $e^+e^-$ cross sections~\cite{DONG20A} and exclusive $e^+e^-$ cross sections to open-bottom~\cite{MIZUK21bb} and closed-bottom~\cite{MIZUK19} final states, we are now in a position 
to perform the first global and unitary analysis of the 
vector bottomonium system above $B\bar{B}$ threshold. 
We use the $K$-matrix formalism for this analysis. In the literature, the $K$-matrix is regularly used in the spectroscopy of hadrons containing $u$, $d$ and $s$ quarks, for example in studies of scalar mesons in Refs.~\cite{albrecht,JPAC,BONN}. A first application of a $K$-matrix in the analysis of heavy quarkonia was made by Uglov \textit{et al.}~\cite{Uglov:2016orr}, performing a unitary coupled channel analysis of $e^+e^-$ annihilation to the $D\bar{D}$, $D^*\bar{D}$ and $D\bar{D}\pi$ channels for energies up to 4.7~GeV.  
Masses and partial decay widths for the charmonium resonances $\psi(3770)$, $\psi(4040)$, $\psi(4160)$ and $\psi(4415)$ were extracted. Comparing to Ref.~\cite{Uglov:2016orr}, our study of the bottomonium system includes three-body final states, allows for non-resonant scattering, is further constrained by the total inclusive cross section and allows for an analytic continuation to the complex energy plane. 

In the following, we describe our model (Sec.~\ref{sec:model}), the datasets used in the analysis (Sec.~\ref{sec:data}), and the fit procedure (Sec.~\ref{sect:fit}). 
In Sec.~\ref{sect:results}, we present our results for the pole positions of the $\Upsilon(4S)$, $\Upsilon(10750)$, $\Upsilon(5S)$, and $\Upsilon(6S)$, as well as estimates for their partial widths to all considered channels.
We discuss the results in Sec.~\ref{sect:int},
comparing them to a variety of theoretical calculations,
and we share our conclusions in Sec.~\ref{sect:concl}. Finally, a more detailed description of the $K$-matrix formalism is given in the Appendix (App.~\ref{appA}) together with a discussion of an analysis that omits  three-body channels (App.~\ref{appB}).


\section{K-matrix Formalism}
\label{sec:model}

Our fit to the data will employ a $K$-matrix that includes resonant and non-resonant scattering terms and is modeled as follows:
\be
K_{\mu,\nu} = \sum_R \frac{g_{R:\mu} g_{R:\nu}}{m_R^2 -s} + f_{\mu,\nu}. \label{eq:ourK}
\ee
The index $R$ refers to resonances while Greek indices denote continuum channels. Except where noted below, the resonant couplings~($g$) and non-resonant terms~($f$) incorporate energy-dependent form factors~(see 
Eqs.~\ref{eq:gFormFactor} and~\ref{eq:fFormFactor})
that are meant to capture the hadronic nature of the relevant interactions. Form factors are parameterized in terms of a universal scale $\beta$, while the resonant and non-resonant coupling strengths are denoted $\hat g$ and $\hat f$, respectively. Derivations and further details can be found in  Appendix \ref{appA}.

The scattering amplitude is written as
\be
\M = (1+KC)^{-1}K
\ee
where $C$ is the Chew-Mandelstam function, with the property 
\be
\Im(C) = - \rho
\ee
and $\rho$ is the diagonal phase space matrix with elements
\be
\rho_{\mu,\nu} = \delta_{\mu,\nu} \frac{k_\nu(s)}{8 \pi S_\nu  \sqrt{s}}.
\ee
Here, $S_\nu$ is a symmetry factor and $k_\nu$ is the center of mass momentum in channel $\nu$. A detailed description of the model choices and our reasons for them is given in  Appendix \ref{appA}.  

Some of the fits described below employ Aitchison's $P$-vector formalism for production~\cite{ian}. In our case this is implemented as
\be
\M_{\mu,ee} = \sum_{\nu}(1+\hat K \hat C)^{-1}_{\mu,\nu}P_{\nu}
\ee
where $P_\nu = K_{\nu,ee}$,
and $\hat K$ and $\hat C$ are defined in the restricted channel space that does not include the initial channel. No form factors were used in the $e^+e^-$ channels of the production vector, as these are not relevant for leptons.

Additional modeling is required since we are fitting  three-body channels in a two-body formalism. We have chosen two approaches:

(i) Create mock two-body channels for $(b\bar b)\pi\pi$ consisting of the bottomonium state and a ``$\pi\pi$" state of mass $2 m_\pi$ with relative angular momentum set to zero. This is a common method used in hadronic  scattering analyses.

(ii) We note that all the three body channels have cross sections of order 10 pb, while the open bottom two-body channels are order 100 pb. Thus it is reasonable to treat the three-body channels perturbatively, neglecting terms of order $g_{R:\textrm{3body}}^2$. This can be realized by placing the production ($e^+e^-$) channel in the $K$-matrix and the three-body channels into a ``final state" matrix, which we call $F$. 
Thus (see Appendix \ref{appA} for details)
\be
\M_{\Delta,ee} = \sum_{\mu} F^{(\Delta)}_\mu (1+\hat C \hat K)^{-1}_{\mu,ee}
\ee
where $\Delta$ denotes a three-body channel, $\hat K$ is defined in the restricted channel space that includes two-body channels and the production channel, and 
\be
F^{(\Delta)}_\mu = K_{\Delta,\mu}.
\ee

We are at liberty to choose an appropriate model for the $F$ vector. 
Two different model choices will be tested: 

(iia) Resonances $R$ feed the three-body channels $ \Delta$, yielding
\be
F^{(\Delta)}_{\mu} = \sum_R\frac{g_{R:\Delta}\, g_{R: \mu}}{m_R^2-s}.
\label{eq:F1}
\ee
(Notice that the units of $g_{R:\mu}$ are dimension one while $g_{R:\Delta}$ are dimension zero.)

(iib) The two-body channel $\mu$ feeds the three-body channel $\Delta$ directly,
\be
F^{(\Delta)}_{\mu} = f_{\Delta:\mu}. 
\label{eq:F2}
\ee

As a further extension, which we did not pursue,
intermediate states saturating
$\pi\pi$ (say an $f_0$) in a three-body channel like $\pi\pi\Upsilon$ could be included
\be
F^{(\Delta = \pi\pi\Upsilon)}_{\mu} = \frac{g_{R:\mu} \cdot g_{R:f_0 \Upsilon} \cdot g_{f_0:\pi\pi}}{(m_R^2-s)\, (m^2_{f_0} - s_{\pi\pi})} \quad  \label{eq:pf3bodyfull}
\ee
if efficiency corrected Dalitz plots were available.

Form factors were not incorporated in the modeling of three body channels since this  was judged to be an unnecessary complication.

Once the scattering amplitude is computed, cross sections are obtained by integrating over the relevant Dalitz plot region. This method comprises a rigorous, if perturbative, approach to solving the three-body problem in the $K$-matrix formalism. The general problem requires solving integral equations and has a long history of investigation~\cite{3bodyK}.
 Results for both of our methods (iia) and (iib) will be reported below.

An additional novel aspect of our approach is the use of the inclusive scattering process. This is achieved by fitting the sum of all channels to $\sigma_{b\bar{b}}=R_{b\bar{b}}\cdot \sigma(e^+e^- \to \mu^+\mu^-)$. An immediate problem that this procedure raises is that the sum over the measured channels falls short of the inclusive cross section for high center-of-mass energies $\sqrt{s}\gtrsim 10.75$ GeV, implying that missing channels can be important. 
Possible missing open bottom channels that are relevant to the energy region of the fit are listed in Table \ref{tab:ch}. Because this represents a substantial source of modeling ignorance, we choose to represent this dynamics with a single dummy channel chosen to correspond to the lightest channel. Thus the channel masses are set to $m(B_s)$ and the relative angular momentum is $\ell =1$. We have tested sensitivity to these choices by adjusting the dummy threshold to 
$m(B) + m(B^\ast) + m(\pi)$ and $2\cdot m(B^\ast) + m(\pi)$ and find negligible difference.

\begin{table}[ht]
\begin{tabular}{c|rc}
\hline\hline
channel & threshold & $\ell$ \\
\hline
$B_sB_s$ & 10734 & 1 \\
$B_sB_s^*$ & 10782 & 1 \\
$BB_1$ & 11000 & 0 \\
$BB_1'$ & $\approx$ 11015 & 0 \\
$B^*B_0$ & $\approx$ 11055 & 0\\
\hline\hline
\end{tabular}
\caption{Some missing two-body channels.}
\label{tab:ch}
\end{table}

Additional error can arise in the model due to neglected resonances with masses much greater than $\sqrt{s}$ in the fit region. If considered, these would contribute
\be
\delta K_{\mu,\nu} \approx \frac{g_{R:\mu} g_{R:\nu}}{m_R^2}
\ee
to the $K$-matrix. This is equivalent to a non-resonant interaction, and is therefore accounted for in the fit to a large extent.


Lastly, we consider neglected high mass channels. In this case the relative channel momentum will be zero and there will be no effect in the scattering amplitude for P-waves and beyond. For S-waves the imaginary part of the amplitude denominator will not change. The real part will shift by $\hat{g}^2[1/(8\pi^2) - s/(96\pi^2M^2)+\ldots]$ and thus will be largely absorbed in the bare parameters, and will not affect the physical properties of the amplitude.

\section{Data Sets} 
\label{sec:data}

With Belle, BaBar, CLEO, and CUSB, four different experiments made contributions to the study of vector-bottomonium states in electron-positron annihilation in the past, with the latter three focused on the inclusive cross section, while the former also studied exclusive cross sections at various center-of-mass energies above the open-bottom threshold.

For this work, we use measurements of the ratio of inclusive bottom anti-bottom quark production over muon production
\begin{equation}
    R_{b\bar{b}} = \frac{\sigma(e^+e^-\to b\bar{b})}{\sigma(e^+e^-\to\mu^+\mu^-)}
\end{equation}
by BaBar and Belle~\cite{AUBERT09E, SANTEL16}. 
This ratio $R_{b\bar{b}}$ is obtained from a measurement of the total hadronic cross section, subtracting $u, d, s, c$-quark contributions extrapolating a model fitted to experimental data of $\sigma(e^+e^-\to\textrm{hadrons})$ below the open-bottom region (see Ref.~\cite{Belle:2006jvm} for details).
A detailed comparison between the BaBar and Belle measurements and a determination of radiative corrections has been performed in Ref.~\cite{DONG20A}. We will use the combined data reported in Ref.~\cite{DONG20A} for the dressed $R_{b\bar{b}}$ ratio and multiply by $\sigma(e^+e^-\to\mu^+\mu^-)=\frac{4\pi\alpha^2}{3s}$ to obtain the total dressed cross section for inclusive bottom-quark pair production $\sigma_{b\bar{b}}$.

In addition, the Belle collaboration recently published cross section measurements of many exclusive open- ($e^+e^-\to (b\bar{q}) (\bar{b}q)$) and hidden-bottom ($e^+e^-\to(b\bar{b})\pi\pi$) processes. The open-bottom processes include $B\bar{B}$, $B^{\ast}\bar{B}$ and $B^\ast\bar{B}^\ast$~\cite{MIZUK21bb} and $B_s^\ast \bar{B}_s^\ast$~\cite{BelleBsBs}. Here and in the following, $B^{(\ast)}_{(s)}\bar{B}^{(\ast)}_{(s)}$ is an abbreviation including charge-conjugated channels as well as the combinations of two neutral and two oppositely charged open-bottom mesons. For the $B\bar{B}$, $B^{\ast}\bar{B}$ and $B^\ast\bar{B}^\ast$ channels dressed cross sections only corrected for ISR effects were published. For $B_s^\ast \bar{B}_s^\ast$, only ``visible" (directly measured) cross sections were made available. 
Initial state radiation effects are in general reaction-dependent and can have an influence on the reconstruction efficiency, so that there is no straightforward way to convert the visible cross section for $e^+e^-\to B_s^\ast \bar{B}_s^\ast$ into a dressed cross section. Instead, we use the visible cross section in our fit for this process alone and thus implicitly assume that the ISR correction is small compared to the sizable uncertainties of the measurement.

The hidden-bottom channels measured by Belle were $\Upsilon(nS)\pi^+\pi^-$ ($n=1,~2,~3$)~\cite{MIZUK19} and $h_b(nP)\pi^+\pi^- $ ($n=1,~2$)~\cite{MIZUK16}. We also consider the measurements of $\Upsilon(1S)\pi^+\pi^-$ and $\Upsilon(2S)\pi^+\pi^-$ at the $\Upsilon(4S)$ resonance~\cite{MIZUK19, Belleon4S}. For these hidden-bottom channels, we abbreviate $\pi^+\pi^-$ as $\pi\pi$ in the following.
In case of all hidden-bottom processes, Born cross sections---observed cross sections corrected for initial-state radiation (ISR) and vacuum polarization (VP) effects---were published. As VP effects are independent of the final state, we multiply the hidden-bottom cross sections by the VP correction factors given in Ref.~\cite{DONG20A} to turn the Born cross sections into dressed cross sections. If no value of the VP correction factor is available at a given center-of-mass energy, we interpolate linearly between two neighbouring points. Given the large density of data for $R_{b\bar{b}}$ and thus the VP correction factor in Ref.~\cite{DONG20A}, we only have to interpolate over small center-of-mass energy regions.
Ref.~\cite{MIZUK19} shows additional data points in the $\Upsilon(10753)$ region obtained using the ISR method ($e^+e^-\to \gamma_{ISR} \Upsilon(nS)\pi\pi$). These are not publicly available because the effective luminosity changes rapidly with center-of-mass energy for these measurements~\cite{MizukPriv}. These data points are thus ignored here.

In principle, as was mentioned in Sec.~\ref{sec:model} (see Eq.~\ref{eq:pf3bodyfull}), our model could accommodate intermediate resonances in three-body processes. However, no information is available on the center-of-mass energy dependence of the $\Upsilon(nS)\pi\pi$ and $h_b(nP)\pi\pi$ Dalitz plots. These would be of particular interest for future high statistics measurements by Belle~II.

\section{Fit and Analysis Strategy}
\label{sect:fit}

We perform a combined least-squares fit to the data using Eqs.~\ref{eq:ourK},~\ref{eq:M},~\ref{eq:HMSXsec} and \ref{eq:HMSXsec2} with four poles and up to eleven channels ($B\bar{B}$, $B^\ast \bar{B}$, $B^\ast\bar{B}^\ast$, ``$B_s \bar{B}_s$", $B_s^\ast\bar{B}_s^\ast$, $\Upsilon(1S)\pi\pi$, $\Upsilon(2S)\pi\pi$, $\Upsilon(3S)\pi\pi$, $h_b(1P)\pi\pi$, $h_b(2P)\pi\pi$ and $e^+e^-$). A fit with only three poles does not yield a satisfactory result.

We define a total of nine different models, using three different form factor scales $\beta = 0.8$, $1.0$, and $1.2$ GeV and three different treatments of the three-body hidden-bottomonium channels. In the first set of models, three-body channels are described in a quasi two-body approach including a $(\pi\pi)$ quasi-particle with mass $m_{\pi\pi}=2m_\pi$. In the other two sets of models, three-body channels are treated perturbatively, using either a resonant or non-resonant coupling as defined in Eqs.~\ref{eq:F1} and~\ref{eq:F2}. Given the lack of data on the center-of-mass energy dependence of the $\Upsilon(nS)\pi\pi$ and $h_b(nP)\pi\pi$ Dalitz plots, we do not use $F^{(\Delta)}_\mu$ given in Eq.~\ref{eq:pf3bodyfull}. 
In addition, we also consider three two-body models applied to the open-bottom channels only as a test of the importance of the three-body data and the robustness of the conclusions (see Appendix \ref{appB}).

The difficulty in estimating initial parameters, the large number of parameters and, by construction, strong correlation between them, forced the adoption of a step-by-step fit procedure. 
We start using a single pole and a single channel ($B \bar{B}$) to describe the total cross section $\sigma_{b\bar{b}}$ in the $\Upsilon(4S)$ region below the $B^\ast \bar{B}$ threshold, where $\sigma_{b\bar{b}}\approx \sigma(e^+e^-\to B\bar{B})$ and $Br(\Upsilon(4S)\to B \bar{B})\approx 100\%$. Then, we add the remaining three poles together with the $B \bar{B}$ data. In an iterative procedure, we then add a new channel to absorb differences between the exclusive and the inclusive cross section, re-fit, and add the exclusive data for that channel. These steps are repeated until all open-bottom channels have been included, with the ``$B_s \bar{B}_s$" channel, used to absorb missing intensity,  added last. For the ``$B_s \bar{B}_s$" channel, we fix the non-diagonal background parameters $\hat{f}_{\mu,\nu}$ ($\mu\neq\nu$) to zero. For all other open-bottom channels, all background terms $\hat{f}_{\mu,\nu}$ are free parameters (see Eq.~\ref{eq:fFormFactor} for the definition of the coupling $\hat f$). 

In addition, the bare masses $m_R$ and couplings $\hat{g}_{R:\mu}$ are free parameters in the fit, apart from the couplings of the lowest mass pole for which we fix $\hat{g}_{R:\mu}=0$ for $\mu \neq B\bar{B}$ owing to the experimental observation that the $\Upsilon(4S)$ is fully saturated by its decay to $B\bar{B}$.

Based on these fits, we add the three-body hidden-bottom processes one-by-one and re-fit the data for each step.
Here, we use the three different approaches of adding either quasi two-body channels or treating the three-body channels perturbatively in one of the final state matrices $F$ detailed above.
In the former case, we fix the background terms in the $K$-matrix $\hat{f}_{\mu,\nu}$ to zero if $\mu$ or $\nu$ is a three-body channel, but allow for a free background term in the production $P$-vector. 

To determine statistical uncertainties of the fit, we use a bootstrapping method, generating 1000 pseudo datasets by randomly varying all data points following Gaussian distributions $G(\sigma,\delta\sigma)$ where $\sigma$ and $\delta\sigma$ are the measured cross section and its uncertainty at any given $s$. Given that Belle did not use a prior requiring $\sigma$ to be positive, we allow random variations to negative cross sections as well.
For each pseudo dataset, we repeat the fit starting from the nominal solution for a given model choice.
To cross check our result, each fit result to a pseudo dataset is tested as a new set of starting parameters in the fit to the actual data, and if that fit yields a better result, we replace the previous solution and repeat the determination of uncertainties with the new fit result.
In each channel, we obtain confidence levels for each center-of-mass energy containing the central 68\% (90\%) of all fit variations for a given model. We remark that the nominal fit result does not necessarily lie in the center of the interval, in large part because the fit result is positive by definition, whereas the  (pseudo-)data are allowed to have negative cross section values.

To extract pole locations, in those cases where the $e^+e^-$ channel is contained in the production vector, we extend $K$ and $\M$ by one dimension using $K_{\mu,e^+e^-}=P_{\mu}$ and $K_{e^+e^-,e^+e^-}=\sum_{R}\frac{g_{R:ee}^2}{m_R^2 - s}$  and systematically scan $\M_{e^+e^-,e^+e^-}(s)$ in the complex plane. We look for poles on the nearest sheets defined such that the sign of the imaginary part of the breakup momentum $k_\mu$ is negative for channels whose thresholds are below ($m_{\textrm{thr}}<\Re(m_{\textrm{pole}})$) and positive for channels whose thresholds are above the pole mass. If a candidate for a pole in the complex plane is found, we repeat the scan around the pole with a finer grid. This procedure is repeated multiple times and only those candidates that can still be identified as poles on a grid with eV binning are examined further.

Once pole candidates are identified, we determine the residues $\textrm{Res}_{\mu}$ using a discrete version of Cauchy's residue theorem (this aids in producing stable results):
\be
\textrm{Res}_{\mu} = \frac{i}{N}\sum_{n=1}^{N}  \mathcal{M}_{\mu,\mu}(s_n)\cdot \frac{\rho_\mu(s_n)}{\sqrt{s_n}} \cdot \left(s_n - m^2_{\textrm{pole}}\right) \quad , \label{eq:residue}
\ee
where the sum over $n$ is chosen such that the discrete values of $s_n$ run along a closed circle around the position of the pole candidate $m^2_{\mathrm{pole}}$ in the complex plane.
These residues are related to the partial width of the relevant resonance for channel $\mu$ via $\Gamma_{R,\mu} = |\textrm{Res}_{\mu}|_R$.

It is possible for ``ghost poles" to appear in the formalism.  This is particularly true for the $\Upsilon(4S)$ region where the fit tends to arrange the background terms such that $[1+KC]$ is not invertible for $s=m^2_{\textrm{ghost pole}}$. 
We identify ghost poles by comparing their total widths as determined by the residues to twice the imaginary part of the pole location. Lack of agreement is evidence for a ghost pole. We also adjust the strengths of the couplings $\hat g$ and $\hat f$ by a factor $\lambda < 1$. As this factor is reduced to zero, poles should approach the bare mass on the real axis,
while ghost poles can exhibit other behavior. We have found that both methods yield valuable information; for example, most ghost poles have total widths that are  $\mathcal{O}(10^{5})$ times too small with respect to the imaginary part of the pole location.

\section{Results}
\label{sect:results}

In this section, we first present fit results for our nine model variations~(Sec.~\ref{sec:fits}).  
We then analyze the structure of the solutions in the complex plane.  
This allows us to determine resonance masses and total widths by extracting pole locations~(Sec.~\ref{sec:poles}) and
to determine branching fractions and partial widths 
by calculating the residues of each 
pole in each channel~(Sec.~\ref{sec:residues}).
As a byproduct, we also plot two-body to two-body hadronic cross sections~(Sec.~\ref{sec:xs}).

\subsection{Fits}
\label{sec:fits}

Fig.~\ref{fig:fitresult1} presents experimental data and fit results for the quasi two-body model with $\beta= 1.0$ GeV for the 10 channels under consideration. The inner~(red) and outer~(green) shaded regions show the central 68\%  and 90\% confidence levels as determined with the bootstrap method described in Section \ref{sect:fit}. As can be seen, the fit is quite good, with a $\chi^2$ per degree of freedom of 1.17. Note, however, that the paucity of data above $\sqrt{s} = 11$ GeV leads to ambiguity in the large energy behavior of the exclusive channels, subject to the constraint provided by the inclusive reaction. Furthermore, the dummy channel (panel (d)) takes up substantial strength, especially near the resonances at (approximately) 10.9 and 11.0 GeV. The question of what this channel represents will be considered below. We remark that the spike seen in the left of panel (a) is the trailing edge of the $\Upsilon(4S)$. 

The fit significantly underestimates the data near 10.65 GeV in the $B^*\bar{B}$ channel (panel (b)). This is caused by the low value of $\sigma_{b\bar{b}}$ in this energy region (see panel (l)), where evidently the inclusive data does not agree with exclusive measurements~\cite{mizukTalk}.

Fit results for all nine models are displayed in Fig.~\ref{fig:fitresult2}. One sees that the fits are quite consistent through the regions in which they are constrained. Many of the channels show the ambiguity in the high energy region mentioned above, which is to be expected due to the absence of exclusive data. Chi-squared values for all the fits are given in Table \ref{tab:chi2}, showing similar fit quality for all the models (with perhaps a slight preference for the three-body non-resonant class of models).

\begin{table}[ht]
\begin{tabular}{l|ccc}
\hline\hline
model & $\chi^2$ & ndf & $\chi^2$/ndf \\
\hline
$\beta=0.8$/ 2-body & 417 & 362 & 1.15 \\
$\beta=1.0$/ 2-body & 423 & 362 & 1.17 \\
$\beta=1.2$/ 2-body & 423 & 362 & 1.17 \\
\hline
$\beta=0.8$/ 3-body/ non-res & 392 & 352 & 1.11\\
$\beta=1.0$/ 3-body/ non-res & 413 & 352 & 1.17\\
$\beta=1.2$/ 3-body/ non-res & 381 & 352 & 1.08\\
\hline
$\beta=0.8$/ 3-body/ resonant & 438 & 367 & 1.19\\
$\beta=1.0$/ 3-body/ resonant & 430 & 367 & 1.17\\
$\beta=1.2$/ 3-body/ resonant & 421 & 367 & 1.15\\
\hline\hline
\end{tabular}
\caption{Global $\chi^2$ values for the different fit models.
The number of degrees of freedom (``ndf'') is 
given by the number of data points (424)
minus the number of free parameters in a model.}
\label{tab:chi2}
\end{table}

\subsection{Poles}
\label{sec:poles}

A summary of the extracted pole positions for each model is presented in Table \ref{tab:poles}. The poles are labelled $\Upsilon(4S)$, $\Upsilon(10750)$, $\Upsilon(5S)$, and $\Upsilon(6S)$ because they clearly reproduce the states reported in the RPP~\cite{pdg}. (The RPP names for these states are $\Upsilon(4S)$, $\Upsilon(10753)$, $\Upsilon(10860)$, and $\Upsilon(11020)$ respectively.) 
We also report 68\% confidence intervals. In a few instances the sheet structure of the corresponding pole is ambiguous. In those cases, multiple pole locations are reported with the sheet being indicated by the signs on the imaginary part of the breakup momentum of the five open-bottom channels (ordered by threshold). Intervals reported in square brackets correspond to alternative solutions found during the bootstrap procedure.

Those $T$-matrix poles that correspond to poles of the $K$-matrix
are displayed in color in Fig.~\ref{fig:poles}.
Each point represents a pole from a fit to the 
bootstrap pseudo-data for different model choices. Poles that we identify as spurious, but which nevertheless have sizable residues, are shown in gray. 
The RPP estimates of mass and width~(corresponding to points $M-i\Gamma/2$) are also shown as stars with error bars. 

All models and bootstrap variations agree quite well for the $\Upsilon(5S)$ and $\Upsilon(6S)$ pole positions, with the $\Upsilon(5S)$ agreeing with the RPP estimate, while our fits obtain a total width for the $\Upsilon(6S)$ that is approximately twice that of the RPP. In contrast, the situation for $\Upsilon(4S)$ and $\Upsilon(10750)$ is less clear. 

\begin{widetext}
$\quad$
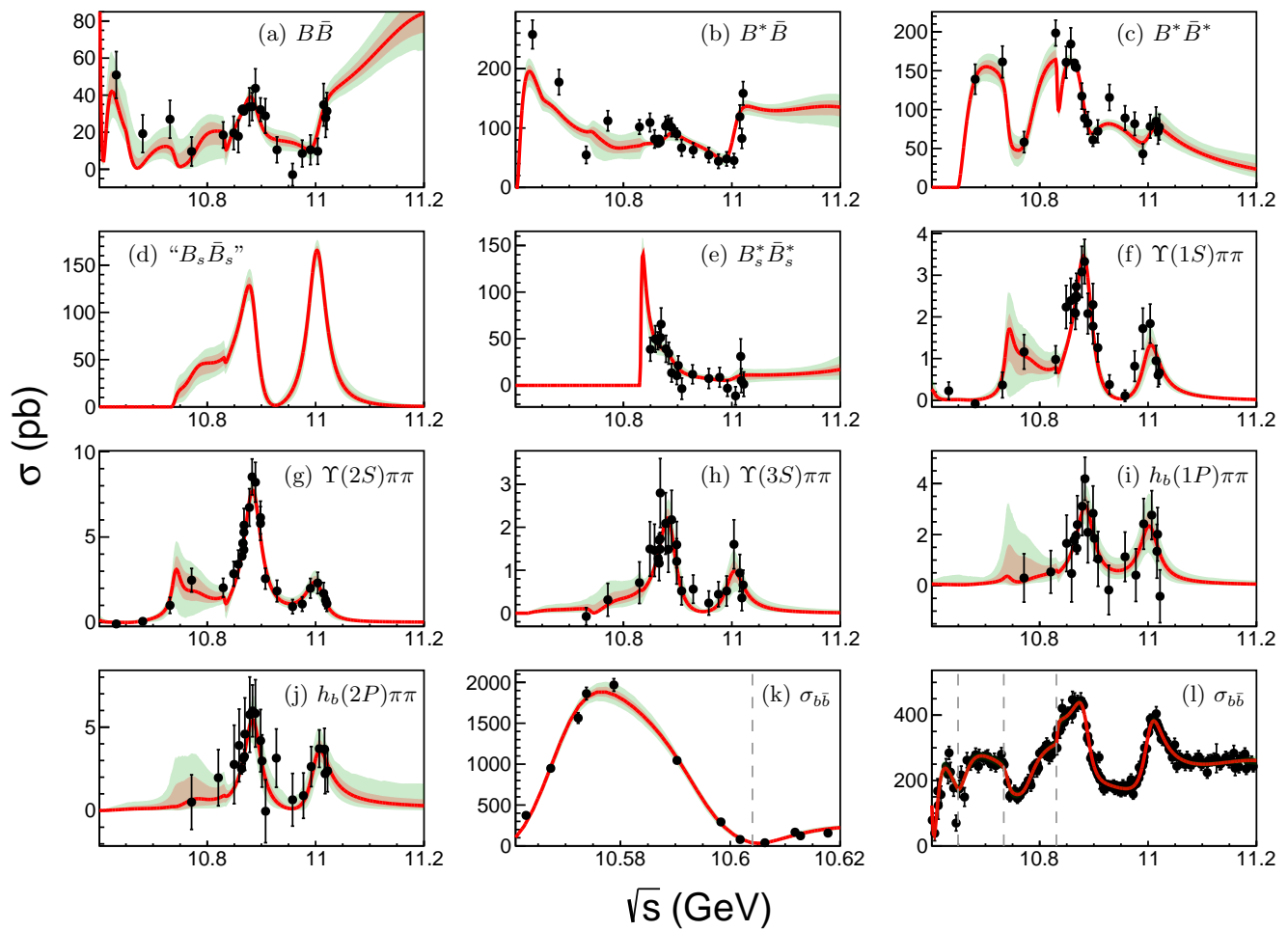
\begin{figure}[ht]
    \centering
    \begin{overpic}[width=1.0\textwidth]{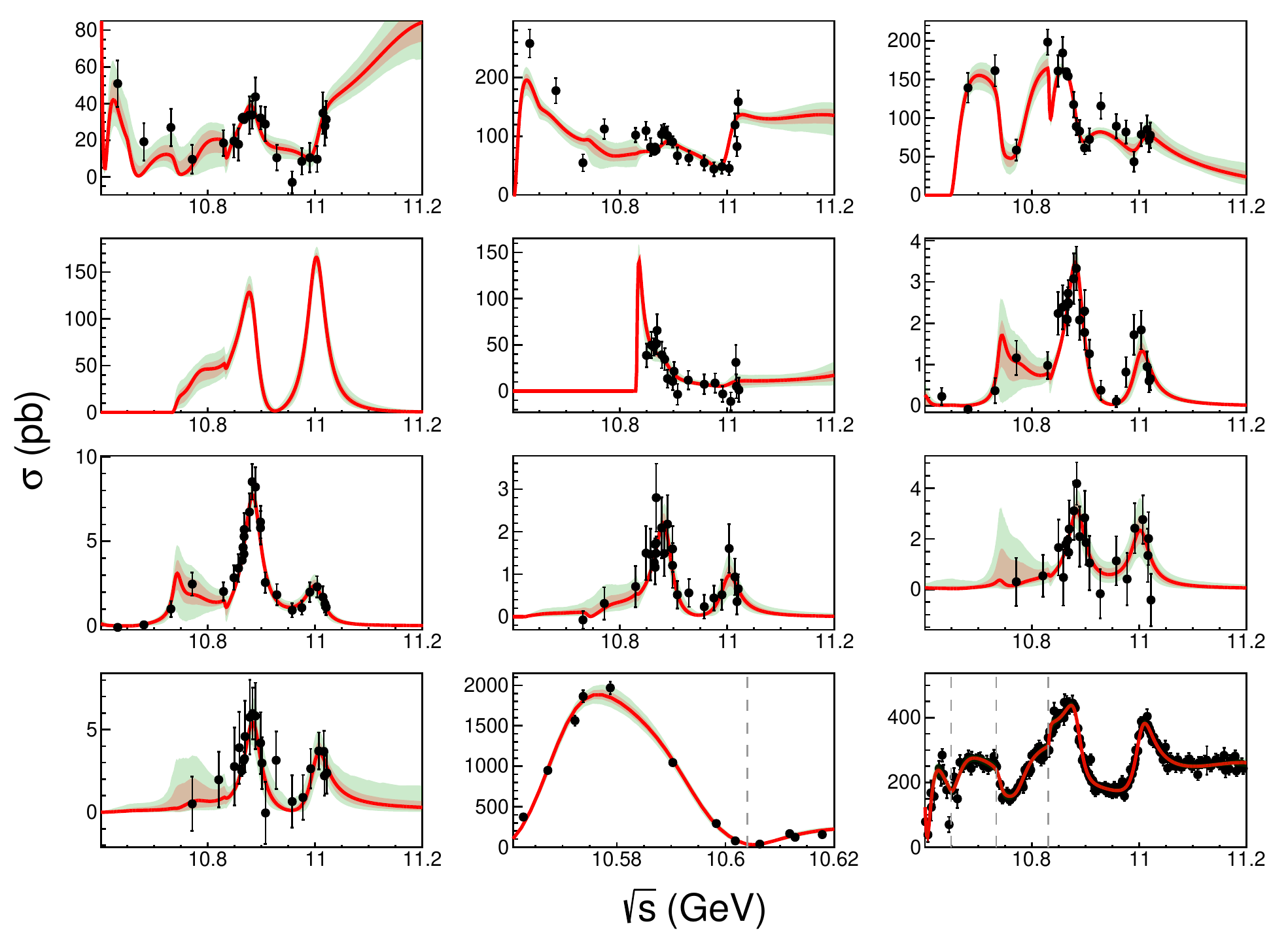}
\put(20,69.5){(a) $B\bar{B}$}
\put(54,69.5){(b) $B^*\bar{B}$}
\put(86,69.5){(c) $B^*\bar{B}^*$}
\put(10,52.5){(d) ``$B_s\bar{B}_s$"}
\put(54,52.5){(e) $B_s^*\bar{B}_s^*$}
\put(86,52.5){(f) $\Upsilon(1S)\pi\pi$}
\put(22,35.5){(g) $\Upsilon(2S)\pi\pi$}
\put(54,35.5){(h) $\Upsilon(3S)\pi\pi$}
\put(86,35.5){(i) $h_b(1P)\pi\pi$}
\put(22,18.7){(j) $h_b(2P)\pi\pi$}
\put(58.7,18.7){(k) $\sigma_{b\bar{b}}$}
\put(91,18.7){(l) $\sigma_{b\bar{b}}$}
    \end{overpic}
    \caption{Fit result for the quasi two-body model 
    with $\beta=1.0~\textrm{GeV}$. Black points are data, the red line is the fit result, and the red and green shaded areas are the central 68\% and 90\% CL regions. The order of the channels is: (a) $B\bar{B}$, (b) $B^*\bar{B}$, (c) $B^*\bar{B}^*$, (d) the dummy channel (``$B_s\bar{B}_s$''), (e) $B_s^*\bar{B}_s^*$, (f) $\Upsilon(1S)\pi\pi$, (g) $\Upsilon(2S)\pi\pi$, (h) $\Upsilon(3S)\pi\pi$, (i) $h_b(1P)\pi\pi$, (j) $h_b(2P)\pi\pi$, (k) $\sigma_{b\bar{b}}$ in the $\Upsilon(4S)$ region, and (l) $\sigma_{b\bar{b}}$ above the $\Upsilon(4S)$ region. The gray dashed lines in (k) and (l) indicate thresholds for $B^*\bar{B}$, $B^*\bar{B}^*$, $B_s\bar{B}_s$ and $B_s^*\bar{B}_s^*$, respectively.}
    \label{fig:fitresult1}
\end{figure}
\end{widetext}

Turning attention to the $\Upsilon(4S)$, we see that most pole positions cluster near the nominal RPP value, although  10-20 MeV higher in mass. There is, however, a region of the $\beta= 1.0$ GeV, three-body non-resonant model poles that lies near $\Im(\sqrt{s}) = -0.03$ GeV. 
The number of data points is sufficiently sparse  (and the model is sufficiently general) 
that perhaps a variety of nearly degenerate minima of the objective function exist. In this case, the secondary group of poles appear to be associated with  $B\bar{B}$ threshold. We note that the secondary group of poles has substantial overlap with a group of ghost poles (indicated in gray). These points tend to move towards the $B\bar{B}$ threshold upon rescaling the couplings -- indicative of their spurious nature, and  hint that the three-body non-resonant secondary group of bootstrap poles should not be considered as viable bottomonium resonance candidates. 

For the $\Upsilon(10750)$, the main accumulation of poles seems to agree with the RPP value.  However, other solutions cannot be ruled out. The high model dependence found in these fits, yielding poles with a large range of masses and widths, reflects the lack of data in the energy region around 10.7 GeV. Some models contained ghost poles in this region that move towards the thresholds of either the dummy or the $B_s^*B_s^*$ channel as the couplings are decreased. New data from an upcoming Belle~II measurement in this energy region will be key in determining the properties of the $\Upsilon(10750)$ with higher precision.

\begin{widetext}
$\quad$
\begin{figure}[ht]
    \centering
    \begin{overpic}[width=1.0\textwidth]{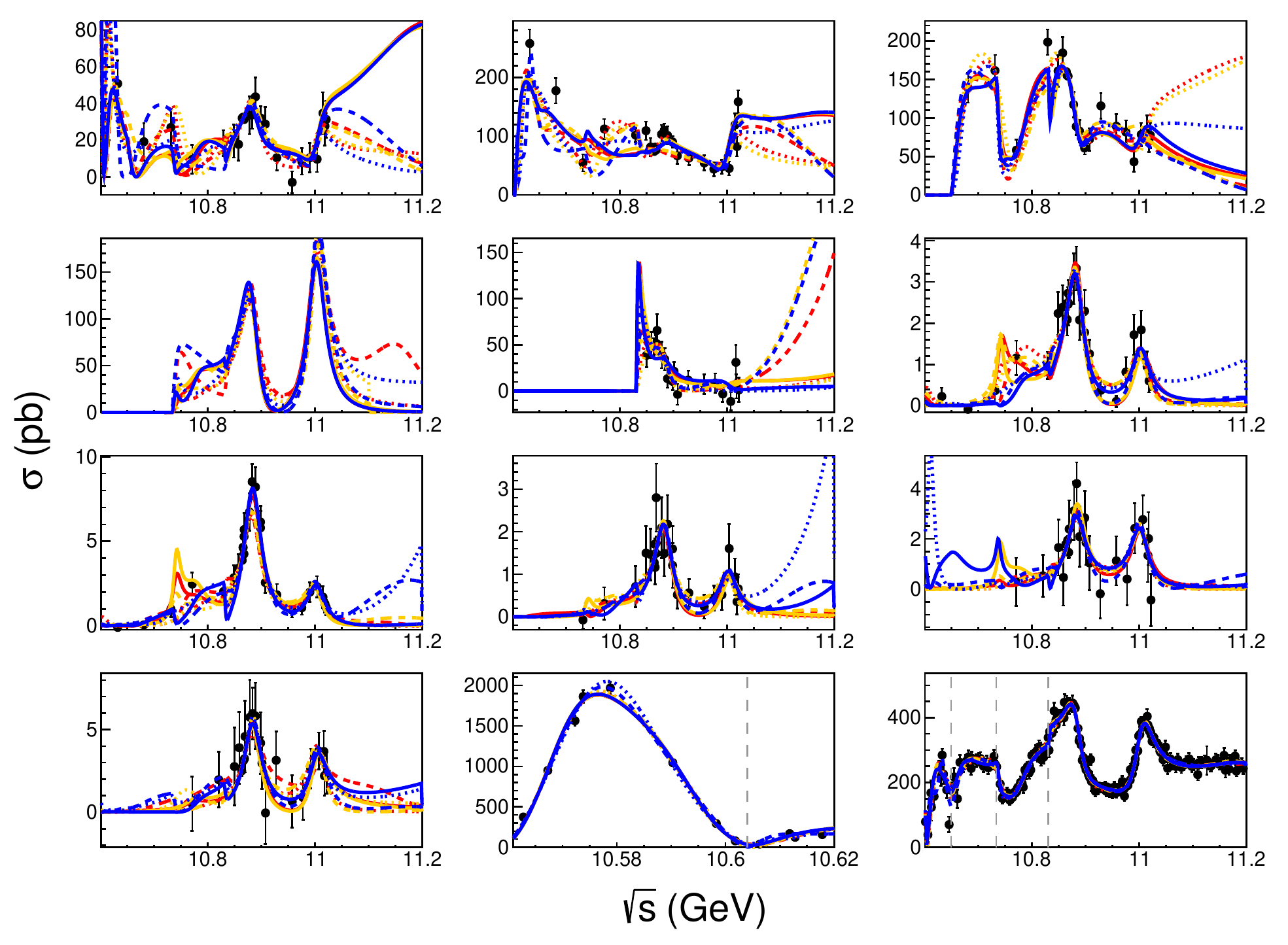}
\put(20,69.5){(a) $B\bar{B}$}
\put(54,69.5){(b) $B^*\bar{B}$}
\put(86,69.5){(c) $B^*\bar{B}^*$}
\put(10,52.5){(d) ``$B_s\bar{B}_s$"}
\put(54,52.5){(e) $B_s^*\bar{B}_s^*$}
\put(86,52.5){(f) $\Upsilon(1S)\pi\pi$}
\put(22,35.5){(g) $\Upsilon(2S)\pi\pi$}
\put(54,35.5){(h) $\Upsilon(3S)\pi\pi$}
\put(86,35.5){(i) $h_b(1P)\pi\pi$}
\put(22,18.7){(j) $h_b(2P)\pi\pi$}
\put(58.7,18.7){(k) $\sigma_{b\bar{b}}$}
\put(91,18.7){(l) $\sigma_{b\bar{b}}$}
    \end{overpic}
    \caption{Comparing fit results for different model choices: Red lines correspond to the quasi two-body models, yellow lines to the three-body models with resonant and yellow lines to three-body models with non-resonant couplings. Solid lines correspond to $\beta=1.0$, whereas dashed and dotted lines correspond to $\beta=0.8$ and $\beta=1.2$, respectively.
    The order of plots (a) through (l) is the same as in Fig.~\ref{fig:fitresult1}.}
    \label{fig:fitresult2}
\end{figure}
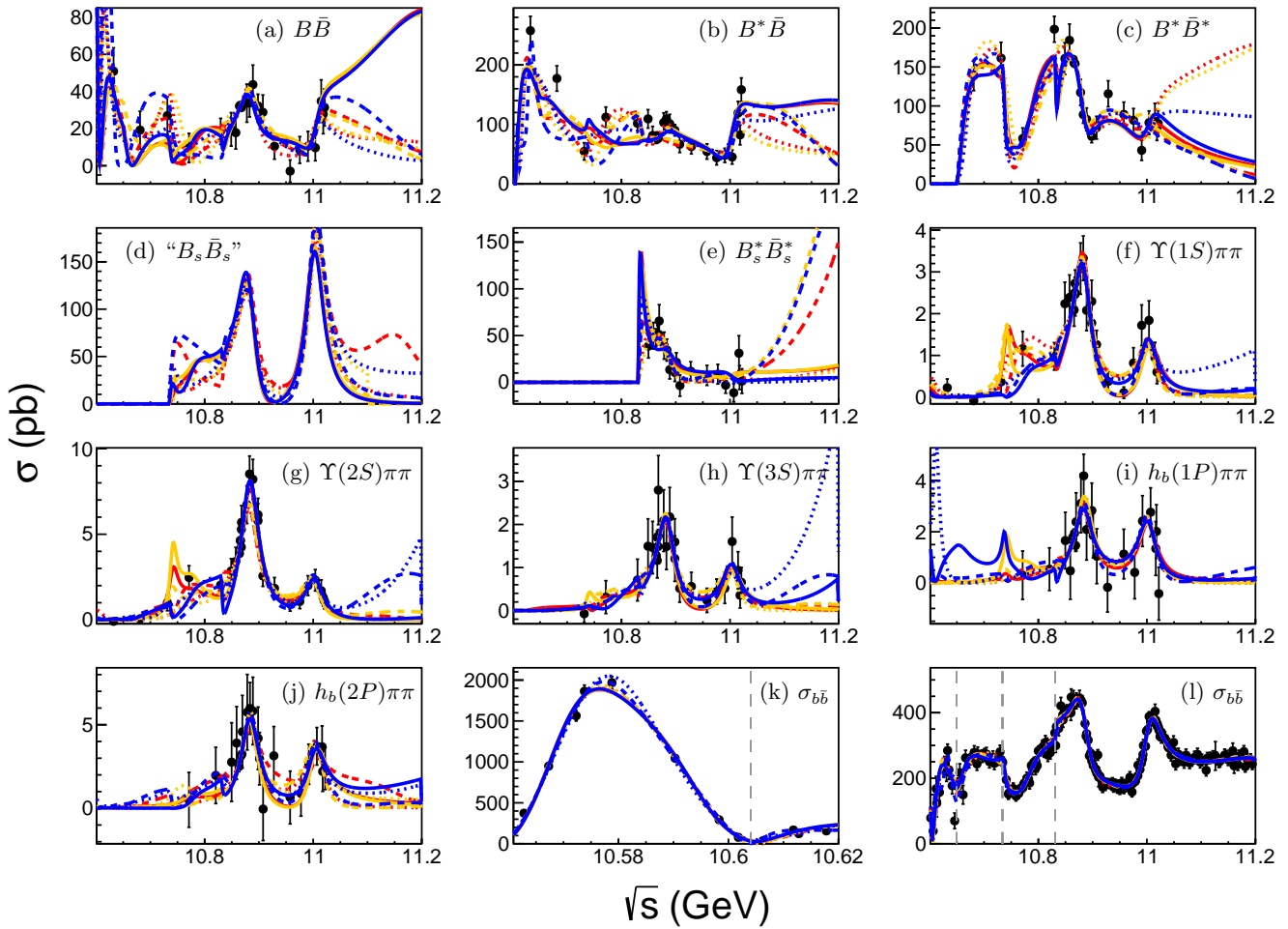
\end{widetext}

\subsection{Residues}
\label{sec:residues}

To determine partial widths and branching fractions, we use the method discussed in Sec.~\ref{sect:fit} to calculate the residues for each channel at each pole in the complex plane.
Considering the significant model dependence and large statistical uncertainties, we do not quote central values, but only estimate ranges for each measurement that 
take into account both the statistical and model spread in our solutions.

\begin{figure}[h!]
     \centering
    \begin{overpic}[width=0.5\textwidth]{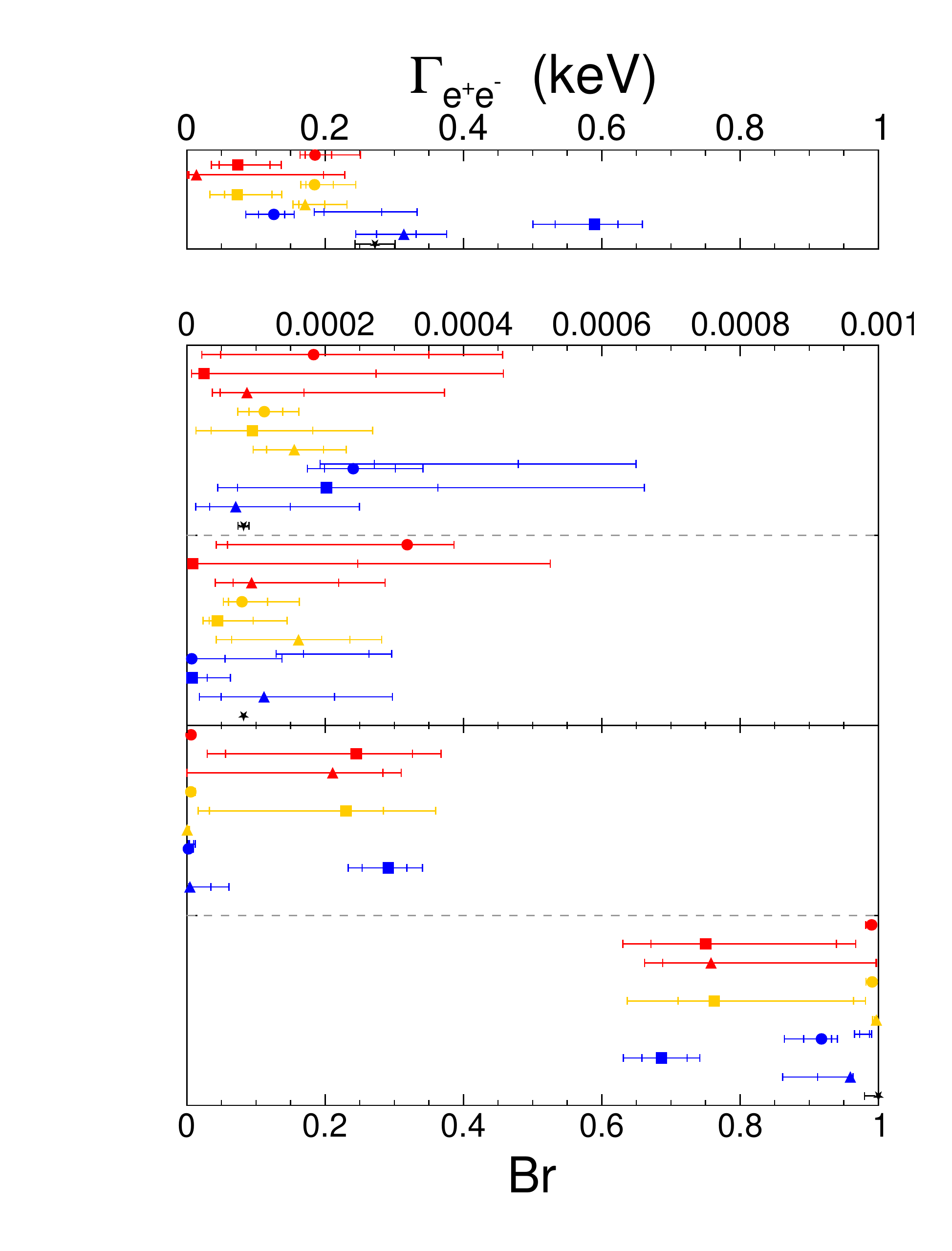}
    \put(7,83.5){ $\mathbf{e^+e^-}$}
    \put(-1.5,63.5){ $\mathbf{\Upsilon(2S)\pi^+\pi^-}$}
    \put(-1.5,48.5){ $\mathbf{\Upsilon(1S)\pi^+\pi^-}$}
    \put(6.5,33.5){ $\mathbf{B^\ast\bar{B}}$}
    \put(7,18.5){ $\mathbf{B\bar{B}}$}
    \end{overpic}
    \caption{Partial widths of the $\Upsilon(4S)$. Red markers correspond to the quasi two-body models, yellow markers to the three-body models with resonant and blue markers to the three-body models with non-resonant couplings. Markers in the same color follow the order $\beta=1.0$ (circles), $\beta=1.2$ (squares), $\beta=0.8$ (triangles). Black stars correspond to the RPP estimate.}
        \label{fig:props4S}
\end{figure}

\begin{figure}[h!]
    \centering
    \begin{overpic}[width=0.5\textwidth]{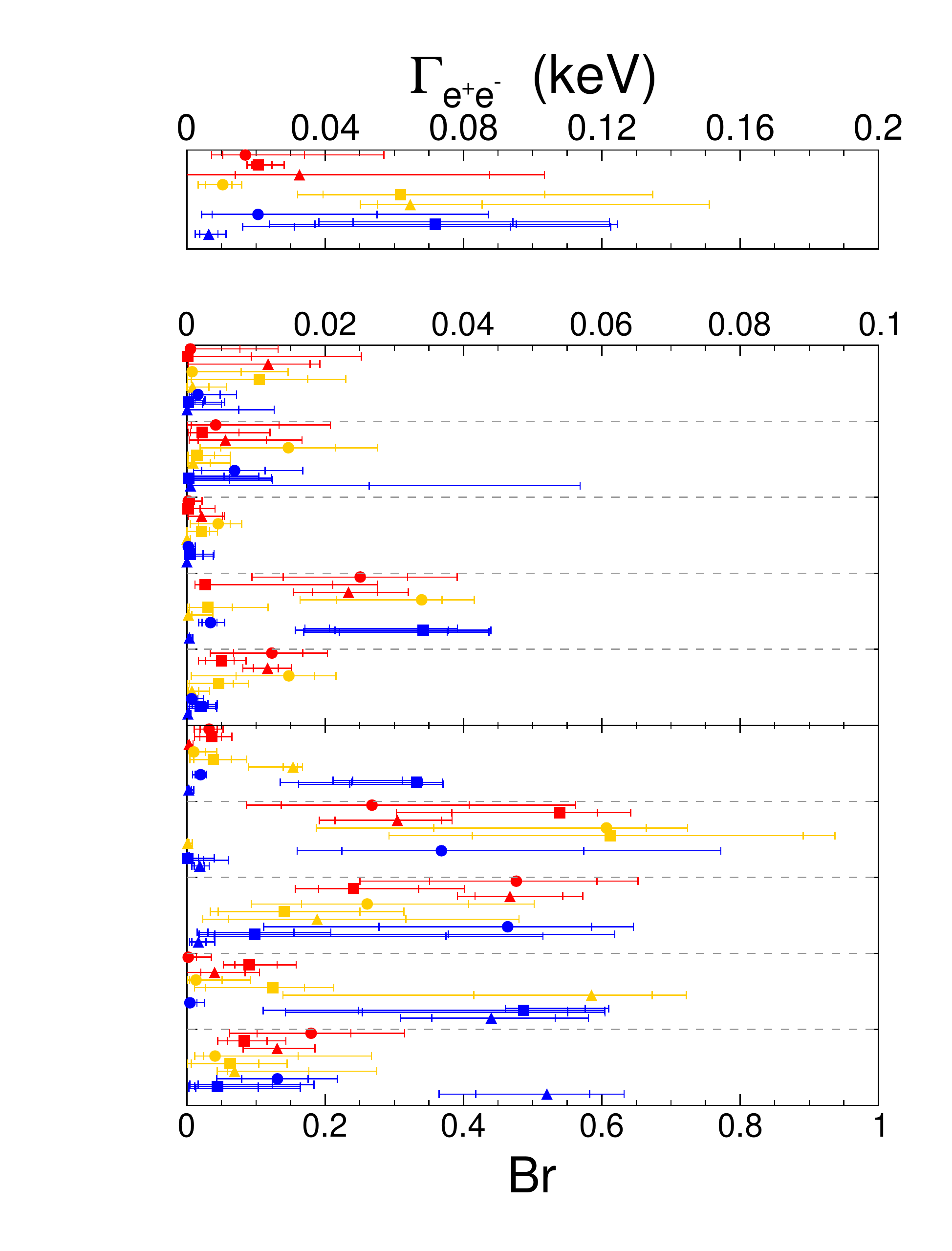}
    \put(6.5,83.5){ $\mathbf{e^+e^-}$}
    \put(-3.0,68){ $\mathbf{h_{b}(2P)\pi^+\pi^-}$}
    \put(-3.0,62){ $\mathbf{h_{b}(1P)\pi^+\pi^-}$}
    \put(-1.8,56){ $\mathbf{\Upsilon(3S)\pi^+\pi^-}$}
    \put(-1.8,50){ $\mathbf{\Upsilon(2S)\pi^+\pi^-}$}
    \put(-1.8,44){ $\mathbf{\Upsilon(1S)\pi^+\pi^-}$}
    \put(4.8,38){ $\mathbf{B^\ast_{s}\bar{B}^\ast_{s}}$}
    \put(3.8,32){ ``$\mathbf{B_{s}\bar{B}_{s}}$"}
    \put(5.5,26){ $\mathbf{B^\ast\bar{B}^\ast}$}
    \put(5.5,20){ $\mathbf{B^\ast\bar{B}}$}
    \put(6.7,14){ $\mathbf{B\bar{B}}$}
    \end{overpic}
    \caption{Partial widths of the $\Upsilon(10750)$, color code is the same as in Fig.~\ref{fig:props4S}.}
    \label{fig:props10750}
\end{figure}

\begin{figure}[h!]
    \centering
    \begin{overpic}[width=0.5\textwidth]{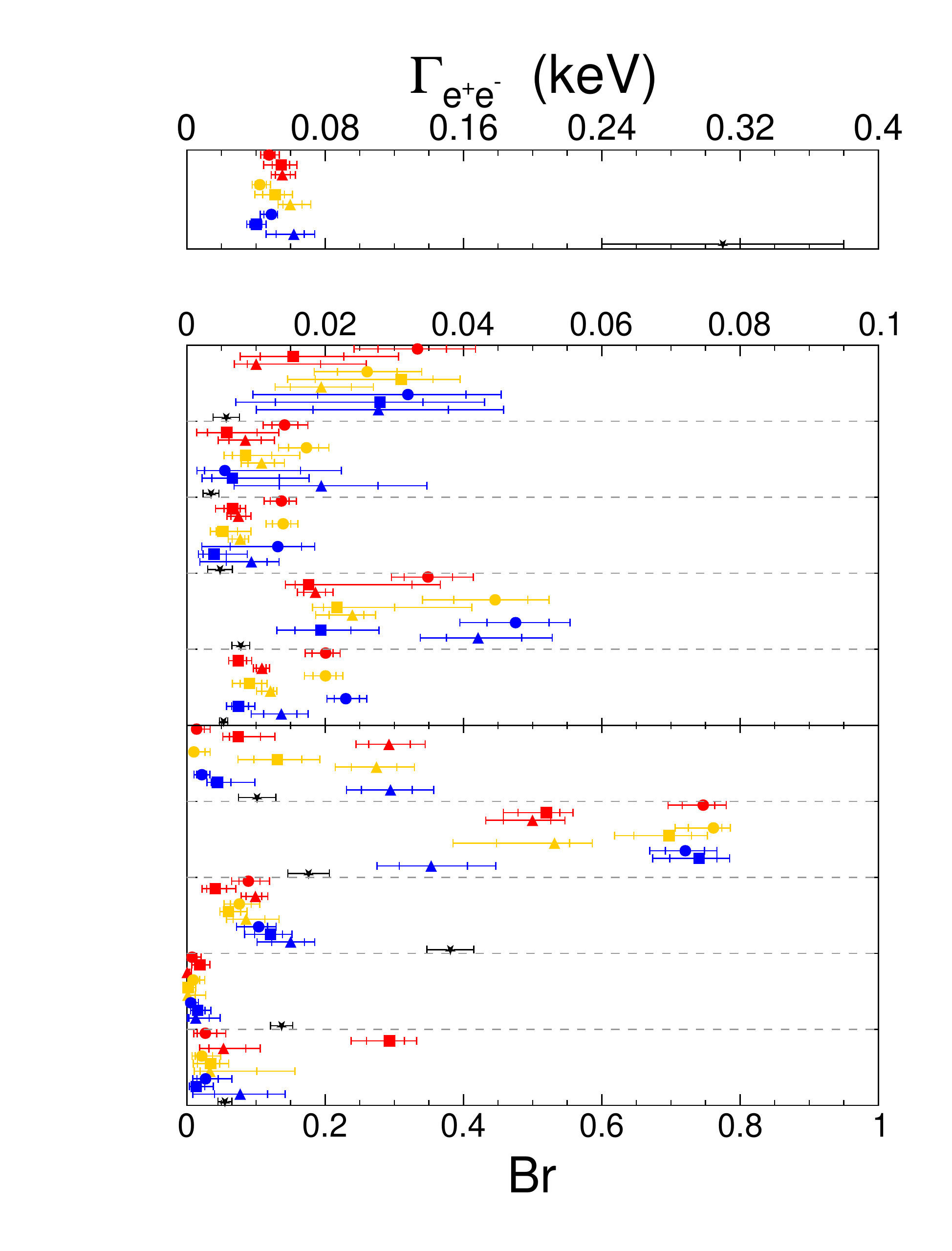}
    \put(6.5,83.5){ $\mathbf{e^+e^-}$}
    \put(-3.0,68){ $\mathbf{h_{b}(2P)\pi^+\pi^-}$}
    \put(-3.0,62){ $\mathbf{h_{b}(1P)\pi^+\pi^-}$}
    \put(-1.8,56){ $\mathbf{\Upsilon(3S)\pi^+\pi^-}$}
    \put(-1.8,50){ $\mathbf{\Upsilon(2S)\pi^+\pi^-}$}
    \put(-1.8,44){ $\mathbf{\Upsilon(1S)\pi^+\pi^-}$}
    \put(4.8,38){ $\mathbf{B^\ast_{s}\bar{B}^\ast_{s}}$}
    \put(3.8,32){ ``$\mathbf{B_{s}\bar{B}_{s}}$"}
    \put(5.5,26){ $\mathbf{B^\ast\bar{B}^\ast}$}
    \put(5.5,20){ $\mathbf{B^\ast\bar{B}}$}
    \put(6.7,14){ $\mathbf{B\bar{B}}$}
    \end{overpic}
    \caption{Partial widths of the $\Upsilon(5S)$, color code is the same as in Fig.~\ref{fig:props4S}.}
    \label{fig:props5S}
\end{figure}

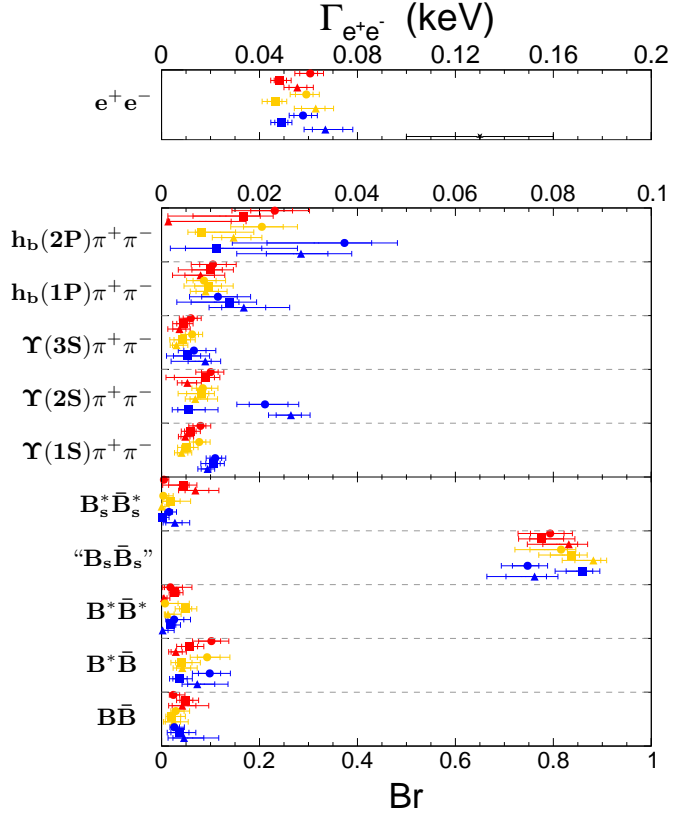
\begin{figure}[h!]
    \centering
    \begin{overpic}[width=0.5\textwidth]{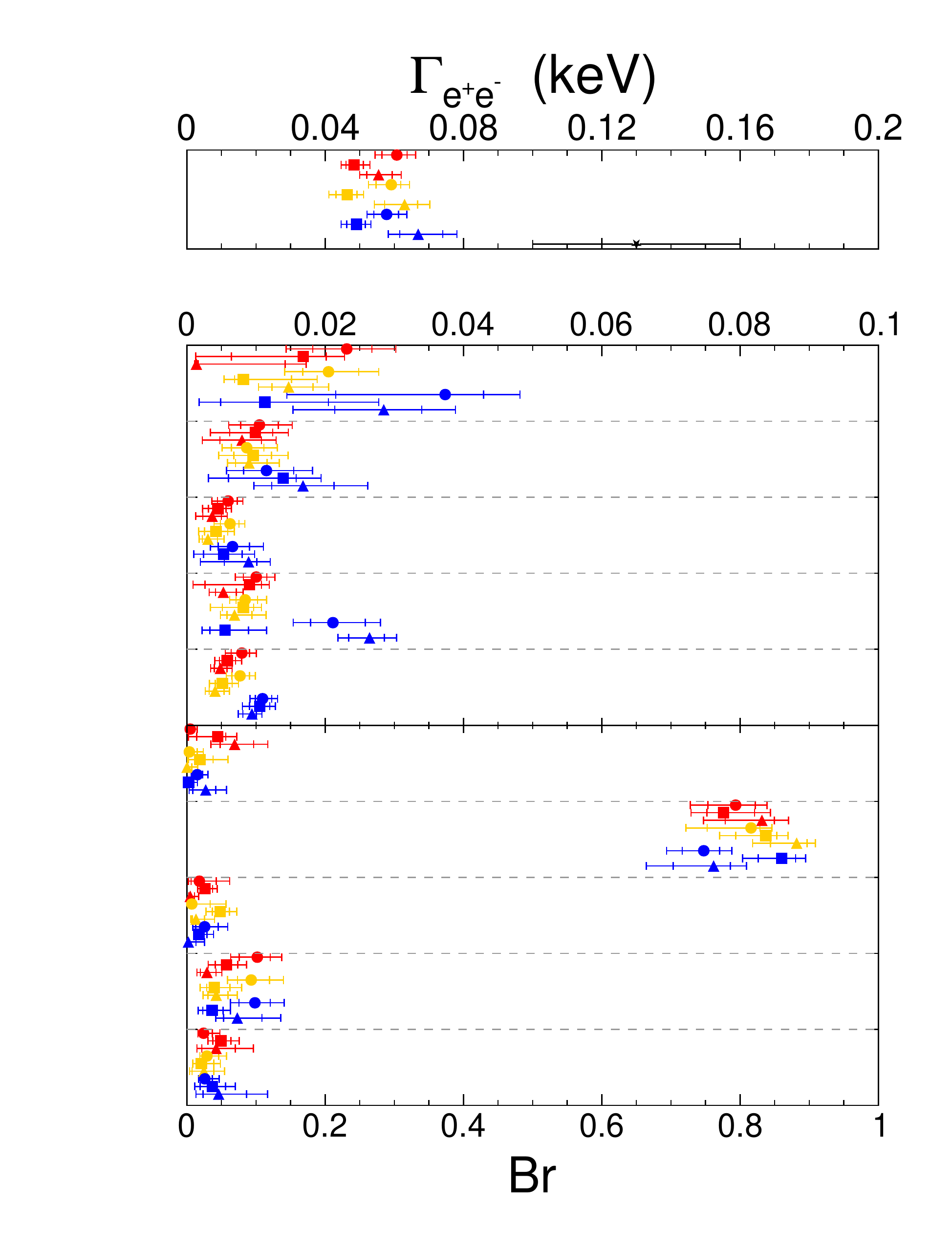}
    \put(6.5,83.5){ $\mathbf{e^+e^-}$}
    \put(-3.0,68){ $\mathbf{h_{b}(2P)\pi^+\pi^-}$}
    \put(-3.0,62){ $\mathbf{h_{b}(1P)\pi^+\pi^-}$}
    \put(-1.8,56){ $\mathbf{\Upsilon(3S)\pi^+\pi^-}$}
    \put(-1.8,50){ $\mathbf{\Upsilon(2S)\pi^+\pi^-}$}
    \put(-1.8,44){ $\mathbf{\Upsilon(1S)\pi^+\pi^-}$}
    \put(4.8,38){ $\mathbf{B^\ast_{s}\bar{B}^\ast_{s}}$}
    \put(3.8,32){ ``$\mathbf{B_{s}\bar{B}_{s}}$"}
    \put(5.5,26){ $\mathbf{B^\ast\bar{B}^\ast}$}
    \put(5.5,20){ $\mathbf{B^\ast\bar{B}}$}
    \put(6.7,14){ $\mathbf{B\bar{B}}$}
    \end{overpic}
    \caption{Partial widths of the $\Upsilon(6S)$, color code is the same as in Fig.~\ref{fig:props4S}.}
    \label{fig:props6S}
\end{figure}

We first consider electronic widths, which are reported in the top panels of Figs. \ref{fig:props4S}, \ref{fig:props10750}, \ref{fig:props5S}, and \ref{fig:props6S} for the $\Upsilon(4S)$, $\Upsilon(10750)$, $\Upsilon(5S)$, and $\Upsilon(6S)$, respectively. 
The results are summarized and compared to theoretical expectations in Tab.~\ref{tab:Gee} of the next section.

Extracted electronic widths for the $\Upsilon(4S)$ state~(Fig.~\ref{fig:props4S}) range from 
0.003 to 0.62~keV,
with most values clustering near 0.15 keV, somewhat lower than the RPP value of  $0.272\pm 0.029$~keV.
A first measurement of the $e^+e^-$ partial width for the $\Upsilon(10750)$ is reported in Fig.~\ref{fig:props10750}, with values ranging from 
0.004 to 0.10~keV.
These values are somewhat smaller than those for the $\Upsilon(4S)$, an issue to which we return in Section \ref{sect:couplings}.
Figures~\ref{fig:props5S} and~\ref{fig:props6S} show that the situation is somewhat cleaner in the cases of the $\Upsilon(5S)$ and $\Upsilon(6S)$, with all extracted partial widths for both
in a range between roughly 0.04 and 0.07~keV.

Our extracted values for both the $\Upsilon(5S)$ and $\Upsilon(6S)$ electronic widths are substantially smaller than the values reported in the RPP, which are $0.31\pm 0.07$ and $0.13\pm 0.03$~keV, respectively.
The original measurements of the inclusive $e^+e^-$ cross sections and their subsequent parameterization~\cite{BESSON85,LOVELOCK85}, which are the basis for the RPP values, were based on very little data and unconstrained models.  
In Ref.~\cite{BESSON85}, for example, the inclusive $e^+e^-$ cross section was modeled using a sum of Gaussian distributions, with a single Gaussian distribution covering the entire $\Upsilon(5S)$ region.
Our model is more fine grained and better constrained by the addition of more experimental data. These circumstances, and the proximity of the recently discovered $\Upsilon(10750)$, drive the large deviations from the RPP values. The implications of this deviation will be explored in Section~\ref{sect:couplings}.

\begin{widetext}
$\quad$
\begin{figure}[h!]
    \centering
    \includegraphics[width=0.8\textwidth]{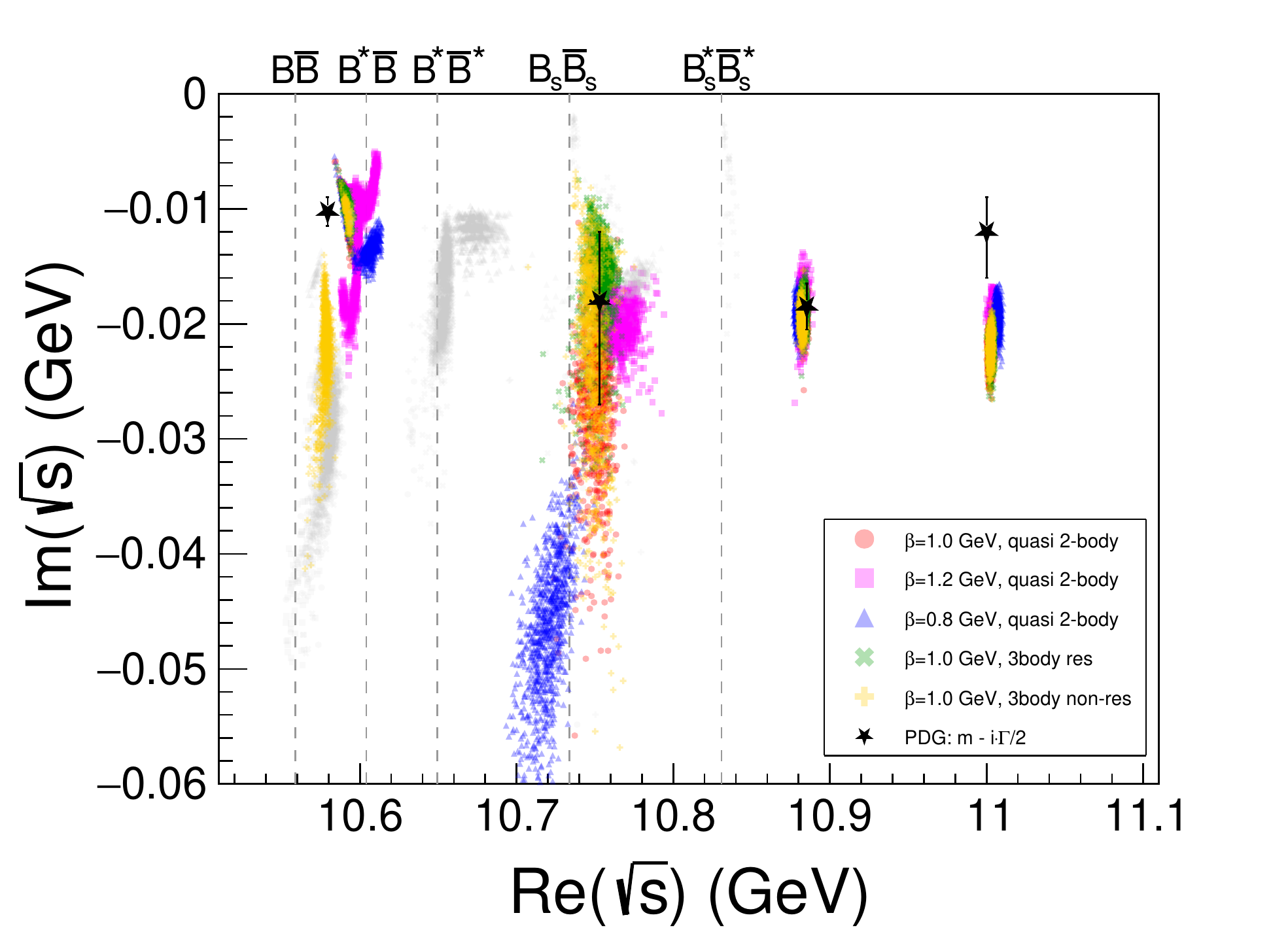}
    \caption{Bootstrap pole positions for five different models indicated by different colors and markers. Gray points with the same marker indicate ghost poles with sizable residues in that model. The black stars represent the RPP estimate using a Breit-Wigner parameterization.} 
    \label{fig:poles}
\end{figure}
\end{widetext}

Branching ratios for hadronic two- and three-body channels are also displayed in Figs.~\ref{fig:props4S} to \ref{fig:props6S} and are summarized and compared to expectations in Tabs.~\ref{tab:4SBR} to \ref{tab:6SBR}. Once again, the model variants permit moderately large  variation in the extracted branching ratios.  This can be due to a number of effects.
For example, we report a nonzero branching fraction of 0.6 (0.4-0.9)\% for $\Upsilon(4S)\to B^*\bar{B}$ for the $\beta=1$ GeV quasi two-body model. This is a result of the pole being found very close to the $B^*\bar{B}$ threshold with a sizable width. In other models, the $\Upsilon(4S)\to B^*\bar{B}$ branching fraction can grow to around 30\%, depending on the location of the pole with respect to the threshold and on the width of the state.

A more subtle example of model variation is visible in Fig. \ref{fig:props5S}, where three branching fractions (circles) for $\Upsilon(1S)\pi\pi$ lie above the other six values. These three points are all models in which $\beta = 1$ GeV, which happen to have a narrower $\Upsilon(10750)$ than the other models, hence less overlap with the $\Upsilon(5S)$, and hence larger $\Upsilon(1S)\pi\pi$ branching fractions.

Branching fractions to the ``missing" channel show some dependence on the state, with values of 
(70-90)\% ($\Upsilon(6S)$), (31-77)\% ($\Upsilon(5S)$), and anywhere between zero and 79\% ($\Upsilon(10750)$). It is perhaps expected that these branching fractions diminish as the mass of the resonance decreases. Nevertheless it is disconcerting that so much of the $\Upsilon(5S)$ and $\Upsilon(6S)$ states disappear into unseen channels. Of course, this may be due to increasing phase space and channels with higher $\sqrt{s}$. It may also be due to the lack of data above 11 GeV. For example, the putative $\Upsilon(6S)$ signal is truncated in the $B^*\bar{B}$ channel by missing data (see panel b of Fig. \ref{fig:fitresult1}), and therefore that 6S branching fraction may be missing some intensity. 

\newpage

\begin{widetext}
$\quad$
\begin{table}[ht]
\footnotesize
\begin{tabular}{l|cccc}
\hline\hline
model & $\Upsilon(4S)$ & $\Upsilon(10750)$ & $\Upsilon(5S)$ & $\Upsilon(6S)$ \\
\hline
RPP (BW)  & $10579.4 \pm 1.2$ & $10753 \pm 6$  & $10885.2 {}^{+2.6}_{-1.6}$ & $11000 \pm 4$ \\
          & $-i(10.25 \pm 1.25)$ & $-i(18 {}^{+9}_{-6})$   &  $-i(18.5 \pm 2)$  & $-i(12 {}^{+4}_{-3})$ \\
 \hline
    $\beta=0.8$/ 2-body&  \textbf{10603}   &   \textbf{10721}   &   \textbf{10880}   & \textbf{11007} \\
                       &  \textbf{$-i$ 14.0}   &  \textbf{$-i$ 45.7}    &  \textbf{$-i$ 19.0}   & \textbf{$-i$ 19.3} \\
(68\% CL)            &  (10590 - 10609)   &   (10711 - 10730)   &   (10879 - 10881)   & (11006 - 11008) \\
                       & $-i$(14.2 - 9.4)    &  $-i$(53.3 - 40.2)    &    $-i$(19.9 - 18.0)    & $-i$(20.4 - 18.3) \\
 $\beta=1.0$/ 2-body&  \textbf{10592}   &  \textbf{10753}    &   \textbf{10884}     & \textbf{11003} \\
  & \textbf{$-i$ 9.4} & \textbf{$-i$ 24.0} & \textbf{$-i$ 19.1} & \textbf{$-i$ 22.5}\\
  (68\% CL) & (10590 - 10593) & (10745 - 10758) & (10883 - 10885) & (11002 - 11003) \\
  & $-i$(10.5 - 8.6) & $-i$(31.3 - 18.5) & $-i$(20.1 - 18.1) & $-i$(23.5 - 21.0) \\
 $\beta=1.2$/ 2-body & \textbf{10606}    &  \textbf{10767}    &  \textbf{10882} & \textbf{11005} \\
    & \textbf{$-i$ 9.5}    &  \textbf{$-i$ 19.7}     &   \textbf{$-i$ 19.0}     & \textbf{$-i$ 19.1} \\
\texttt{(--+++)} & \textbf{10599} & & & \\
 & \textbf{$-i$ 13.1} & & & \\
  (68\% CL)  &  (10599 - 10609)   &   (10762 - 10774)   &  (10881 - 10883)      & (11004 - 11005) \\
   &  $-i$(10.0 - 7.6)  &   $-i$(21.4 - 18.2)   &   $-i$(20.3 - 17.3)     &  $-i$(20.0 - 18.2)\\
\texttt{(--+++)} & (10593 - 10599) & & & \\
& $-i$(18.4 - 10.0) & & & \\
\hline 
 $\beta=0.8$/ 3-body/ non-res &  \textbf{10582}   &   \textbf{10635}   &   \textbf{10881}     & \textbf{11003}\\
                              &  \textbf{$-i$ 18.8}   &  \textbf{$-i$ 7.9}    &   \textbf{$-i$ 21.3}     & \textbf{$-i$ 25.8}\\
        (68\% CL)             &  (10581 - 10583)   &  (10633 - 10638)    &   (10880 - 10883)     & (11002 - 11004)\\
                              &  $-i$(21.2 - 15.8)   &  $-i$(9.1 - 5.7)    &  $-i$(22.3 - 19.3)      & $-i$(27.8 - 23.9)\\
 $\beta=1.0$/ 3-body/ non-res & \textbf{10579}    &   \textbf{10747}   &     \textbf{10882}   & \textbf{11003} \\
                              &  \textbf{$-i$ 21.7}   &   \textbf{$-i$ 20.3}   &   \textbf{$-i$ 20.1}     & \textbf{$-i$ 22.1}\\
        (68\% CL)             &  (10577 - 10580) [(10590 - 10593)]  &   (10743 - 10755)   &   (10881 - 10883)     & (11002 - 11004)\\
                              &  $-i$(25.9 - 18.9) [$-i$(11.4 - 9.4)]  & $-i$(29.8 - 14.7)     &  $-i$(20.6 - 18.7)      & $-i$(23.3 - 20.8)\\
 $\beta=1.2$/ 3-body/ non-res &  \textbf{10578}   &  \textbf{10639}    &   \textbf{10882}     & \textbf{11001}\\
                              &  \textbf{$-i$ 18.9}   &   \textbf{$-i$ 68.0}   &  \textbf{$-i$ 18.4}      & \textbf{$-i$ 19.0}\\
      \texttt{(---++)}               &     &   \textbf{10657}   &        & \\
                              &     & \textbf{$-i$ 62.9}     &        & \\
       \texttt{(----+)}              &     &   \textbf{10657}   &        & \\
                              &     &  \textbf{$-i$ 62.8}    &        & \\
        (68\% CL)             &  (10575 - 10580) [(10598 - 10607)]   &  (10607 - 10658) &   (10881 - 10883)     & (11000 - 11002)\\
                              &  $-i$(22.0 - 16.4) [$-i$(9.0 - 6.3)]  &  $-i$(86.3 - 58.7) &   $-i$(19.5 - 17.2)     & $-i$(19.8 - 18.0) \\
        \texttt{(---++)}                       &     &  (10626 - 10717)    &        & \\
                              &     &  $-i$(78.0 - 48.6)    &        & \\
        \texttt{(----+)}                 &     &  (10626 - 10714)    &        & \\
                              &     &  $-i$(78.3 - 49.3)    &        & \\
\hline
 $\beta=0.8$/ 3-body/ resonant &  \textbf{10592}   &  \textbf{10675}    &   \textbf{10879}     &  \textbf{11003}\\
                               &  \textbf{$-i$ 9.8}   &   \textbf{$-i$ 15.7}   &   \textbf{$-i$ 21.2}     & \textbf{$-i$ 23.9}\\
(68\% CL)                      &  (10591 - 10593)   &  (10663 - 10688)    &  (10878 - 10880)      & (11002 - 11004) \\
                               &  $-i$(10.8 - 9.0)   &  $-i$(17.6 - 14.8)    &    $-i$(22.7 - 20.1)    & $-i$(25.3 - 21.8) \\
 $\beta=1.0$/ 3-body/ resonant &  \textbf{10591}   &  \textbf{10755}    & \textbf{10883}    & \textbf{11003}\\
                               &  \textbf{$-i$ 9.3}   &  \textbf{$-i$ 15.2}    &    \textbf{$-i$ 18.4}    & \textbf{$-i$ 22.4}\\
(68\% CL)                      &  (10.590 - 10.593)   &  (10748 - 10759)    &    (10882 - 10884)    & (11002 - 11004)\\
                               & $-i$(10.4 - 8.7)    &  $-i$(19.2 - 13.1)    &   $-i$(19.6 - 17.4)     & $-i$(23.4 - 20.9)\\
 $\beta=1.2$/ 3-body/ resonant &   \textbf{10606}  &  \textbf{10776}    &  \textbf{10881}      & \textbf{11003}\\
                               &  \textbf{$-i$ 9.7}   &   \textbf{$-i$ 39.8}   &   \textbf{$-i$ 18.7}     & \textbf{$-i$ 18.6}\\
(68\% CL)                      & (10597 - 10607)    &  (10704 - 10781)    &   (10880 - 10883)     & (11001 - 11003)\\
                               &  $-i$(9.9 - 8.1)   &   $-i$(134.3 - 27.0)   &  $-i$(19.8 - 17.1)      & $-i$(19.6 - 17.6)\\
 \hline
 final estimate & (10590 - 10610) & (10630 - 10780)  & (10878 - 10884)  & (11000 - 11008) \\
  & $-i$(6 - 16)  & $-i$(10 - 70)  & $-i$(17 - 22)  & $-i$(18 - 27) \\
\hline\hline
\end{tabular}
\caption{
Pole positions in MeV. Parentheses indicate central 68\% CL regions, square brackets indicate regions of a second, separate bootstrap solution where present. In case the sheet structure of a pole is ambiguous (for broad poles or poles close to a threshold), values extracted from deeper sheets are reported as well. These additional sheets are labelled by``+" and ``-" signs corresponding to the sign of the imaginary part of the breakup momentum of the open-bottom channels, sorted by ascending thresholds.  } 
\label{tab:poles}
\end{table}
\end{widetext}

Lastly, we note that both the $\Upsilon(5S)$ and $\Upsilon(6S)$ are reported to decay to $Z_b\pi$ and $Z^\prime_b\pi$, with the two $Z_b$ states decaying to $B^\ast\bar{B}$ and $B^\ast\bar{B}^\ast$, respectively. This implies that $B^\ast\bar{B}\pi$ and $B^\ast\bar{B}^\ast\pi$ could be important components of the ``missing" channel.

\subsection{Hadronic Scattering Cross Sections}
\label{sec:xs}

With the $K$-matrix in hand it is a simple matter to obtain cross sections for all channels. Here we examine the set of 25 two-body channels to confirm that the model is behaving in a reasonable way. 
The result for the two-body variant with $\beta=1.0$ GeV is displayed in Fig.~\ref{fig:bbtobb} along with the central 68\% and 90\% intervals that are obtained from fits to the pseudo-data samples. The cross sections obtained from all the fit models are shown in Fig.~\ref{fig:bbtobb2}.

We see that the cross sections have the scale of typical hadronic cross sections, which is reassuring. All cross sections also display a sharp threshold rise, which is not unexpected. This is typically followed by a rapid drop over a range of $\sqrt{s} \approx 100$ MeV. Again, this is expected~\cite{Hilbert:2007hc} as this scale is set by the relevant hadronic wavefunctions. Beyond the threshold region there is scant evidence of the bottomonium resonances, with small peaks visible at $\Upsilon(10750)$, $\Upsilon(5S)$ and $\Upsilon(6S)$ in some channels. This is likely due to the large background scattering present and the large couplings to the dummy channel.

It is interesting that the range of fits in Fig. \ref{fig:bbtobb} tend to follow those of Fig. \ref{fig:bbtobb2}. Evidently the different fit models tend to span a similar region in parameter space as the minima obtained in the variant model space. This is a strong indication that the modeling is consistent and that the fluctuations present are due to data quality. 

\section{Discussion and Interpretation}
\label{sect:int}

The large bottom quark mass makes bottomonium an ideal system for the application of constituent quark models. It is therefore informative to compare the results we find to model predictions and lattice field theory computations where they exist.  

\subsection{General Features}

Fig. \ref{fig:fitresult1} reveals that most fit variability lies in the region above $\sqrt{s} \approx 11$ GeV and around $\sqrt{s} = 10.75$ GeV. The former is clearly because of the lack of data in exclusive channels for large energy. As a result the strength in the inclusive measurement can be shared in different ways throughout the available channels. This observation is reinforced by the results of Fig. \ref{fig:fitresult2}, which show that the large energy ambiguity is reflected in the model variant optimal fits. Evidently, model variation  recapitulates data variance, which is a strong indication of the consistency of our approach. As we have noted, the paucity of data above 11 GeV makes it difficult to pin down branching fractions for the $\Upsilon(6S)$; hence measurements in this region would be most welcome.

The other region of greatest model variation occurs near $\sqrt{s} = 10.75$ GeV. As might be guessed, this is correlated with ambiguity in the pole location of the $\Upsilon(10750)$. 
We find very strong evidence for this state with a significance  well above 10 $\sigma$ compared to models using only three poles, where the significance is estimated using the difference in the minimum $\chi^2$ of the respective fits.
However, its location varies with model and under bootstrapping.
The imaginary part of the pole runs between 10 and 70 MeV (with the largest cluster around 20 MeV). More importantly, the real part of the pole lies above or below the nominal missing channel threshold, which leads to a significant spread of the measured branching fractions.

The $\Upsilon(10750)$ was discovered in three-body decays~\cite{MIZUK19} to $\Upsilon(nS)\pi\pi$. Because of this we thought it prudent to determine if evidence for this resonance exists in the two-body open bottom channels. This investigation is reported in Appendix \ref{appB}. We again see strong evidence for the $\Upsilon(10750)$ (similar to Ref.~\cite{DONG20A}).

Panel (d) of Fig. \ref{fig:fitresult2} shows the rate for $e^+e^-$ to the missing dummy channel. One sees that the reaction is dominated by the $\Upsilon(5S)$ and $\Upsilon(6S)$, with a more ambiguous contribution from the $\Upsilon(10750)$. This result is  driven by the lack of agreement between $\sigma_{b\bar{b}}$ (panels k and l) and the sum over the measured exclusive channels. As we have discussed above, we interpret the discrepancy as being due to missing multipion final states such as $B^*\bar{B}\pi$ and $B^*\bar{B}^*\pi$, including $Z_b\pi$ and $Z_b^\prime\pi$ contributions.

\begin{widetext}
$\quad$
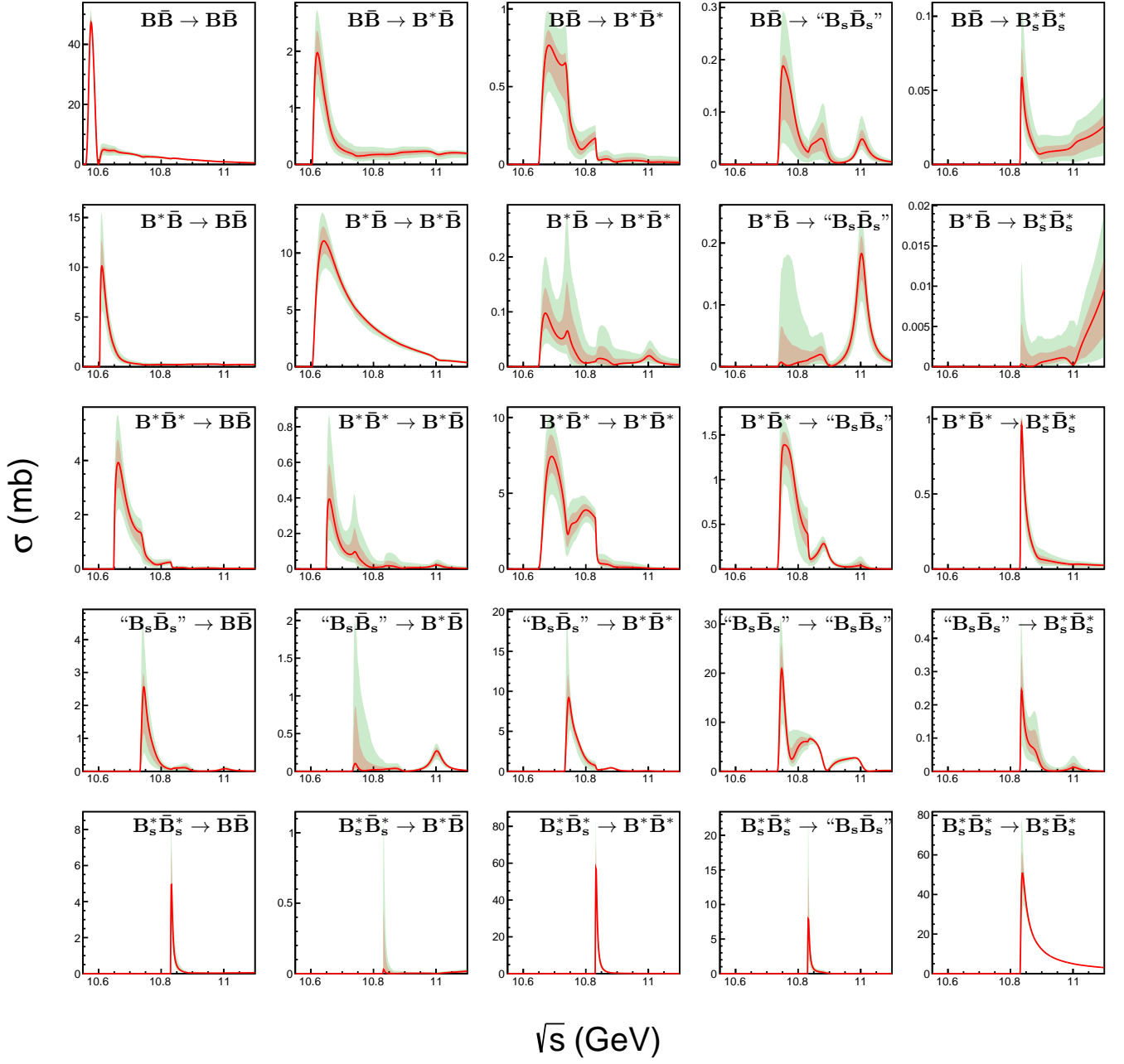
\begin{figure}[h!]
    \centering
    \begin{overpic}[width=1.0\textwidth]{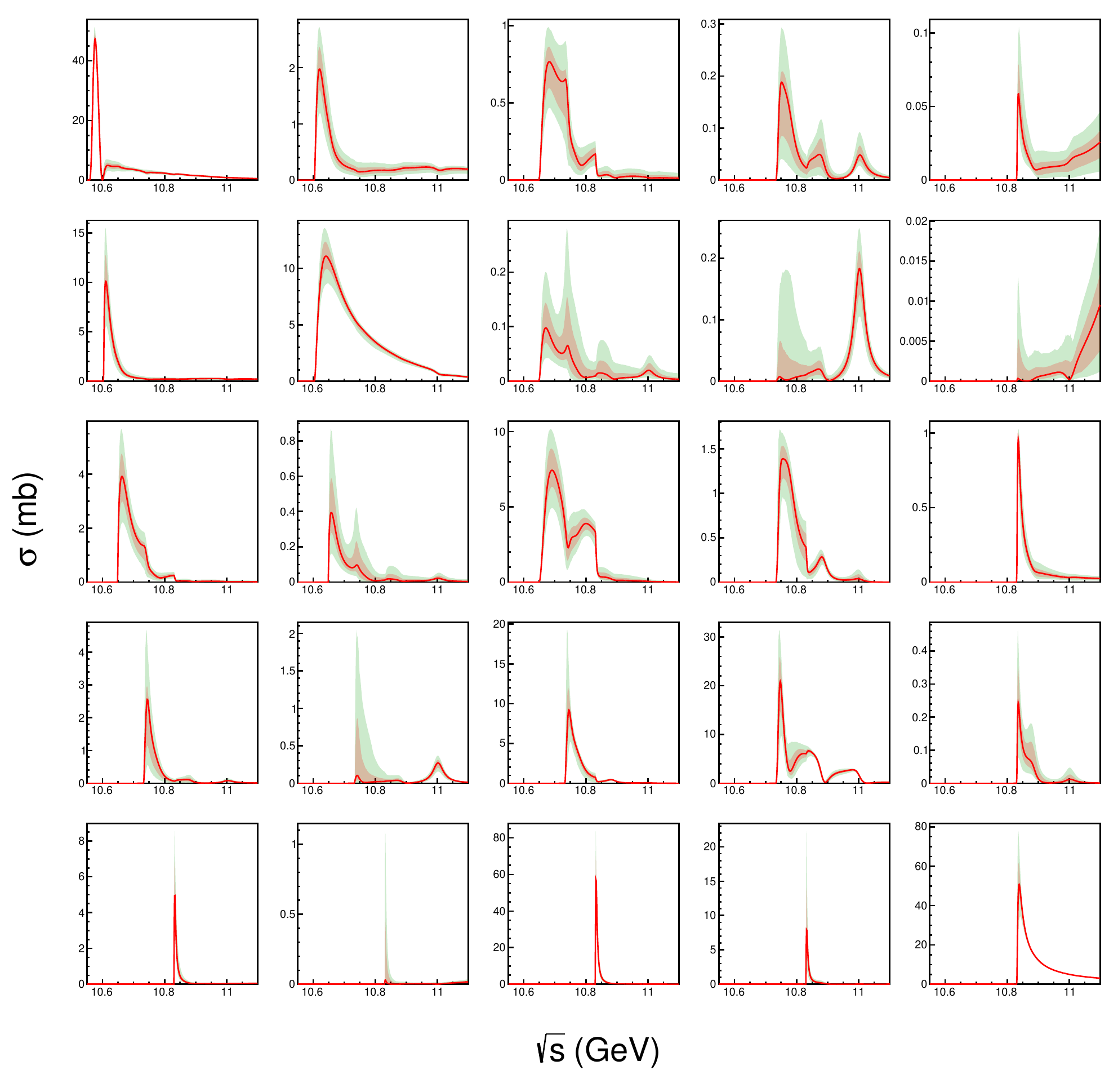}
    \put(13.0,94.5){$\mathbf{B\bar{B}\to B\bar{B}}$}
    \put(13.0,76.5){$\mathbf{B^\ast\bar{B}\to B\bar{B}}$}
    \put(12.5,58.5){$\mathbf{B^\ast\bar{B}^\ast\to B\bar{B}}$}
    \put(11.0,40.5){$\mathbf{``B_s\bar{B}_s"\to B\bar{B}}$}
    \put(12.3,22.4){$\mathbf{B^\ast_s\bar{B}^\ast_s\to B\bar{B}}$}
    
    \put(31.5,94.5){$\mathbf{B\bar{B}\to B^\ast\bar{B}}$}
    \put(31.3,76.5){$\mathbf{B^\ast\bar{B}\to B^\ast\bar{B}}$}
    \put(30.7,58.5){$\mathbf{B^\ast\bar{B}^\ast\to B^\ast\bar{B}}$}
    \put(29.0,40.5){$\mathbf{``B_s\bar{B}_s"\to B^\ast\bar{B}}$}
    \put(30.7,22.4){$\mathbf{B^\ast_s\bar{B}^\ast_s\to B^\ast\bar{B}}$}
    
    \put(49.5,94.5){$\mathbf{B\bar{B}\to B^\ast\bar{B}^\ast}$}
    \put(49.3,76.5){$\mathbf{B^\ast\bar{B}\to B^\ast\bar{B}^\ast}$}
    \put(48.7,58.5){$\mathbf{B^\ast\bar{B}^\ast\to B^\ast\bar{B}^\ast}$}
    \put(47.0,40.5){$\mathbf{``B_s\bar{B}_s"\to B^\ast\bar{B}^\ast}$}
    \put(48.7,22.4){$\mathbf{B^\ast_s\bar{B}^\ast_s\to B^\ast\bar{B}^\ast}$}
    
    \put(67.5,94.5){$\mathbf{B\bar{B}\to ``B_s\bar{B}_s"}$}
    \put(67.3,76.5){$\mathbf{B^\ast\bar{B}\to ``B_s\bar{B}_s"}$}
    \put(66.7,58.5){$\mathbf{B^\ast\bar{B}^\ast\to ``B_s\bar{B}_s"}$}
    \put(65.0,40.5){$\mathbf{``B_s\bar{B}_s"\to ``B_s\bar{B}_s"}$}
    \put(66.7,22.4){$\mathbf{B^\ast_s\bar{B}^\ast_s\to ``B_s\bar{B}_s"}$}
    
    \put(85.5,94.5){$\mathbf{B\bar{B}\to B^\ast_s\bar{B}^\ast_s}$}
    \put(85.3,76.5){$\mathbf{B^\ast\bar{B}\to B^\ast_s\bar{B}^\ast_s}$}
    \put(84.7,58.5){$\mathbf{B^\ast\bar{B}^\ast\to B^\ast_s\bar{B}^\ast_s}$}
    \put(84.7,40.5){$\mathbf{``B_s\bar{B}_s"\to B^\ast_s\bar{B}^\ast_s}$}
    \put(84.7,22.4){$\mathbf{B^\ast_s\bar{B}^\ast_s\to B^\ast_s\bar{B}^\ast_s}$}
    
    \end{overpic}
    \caption{Cross sections for $B_{(s)}^{(\ast)}\bar{B}_{(s)}^{(\ast)} \to  B_{(s)}^{(\ast)}\bar{B}_{(s)}^{(\ast)}$ scattering. The red line shows the fit result for the two-body $\beta = 1.0$ fit while red and green shaded areas indicate the regions containing 68\% and 90\% of the pseudo-data fits.}
    \label{fig:bbtobb}
\end{figure}
\end{widetext}

\begin{widetext}
$\quad$
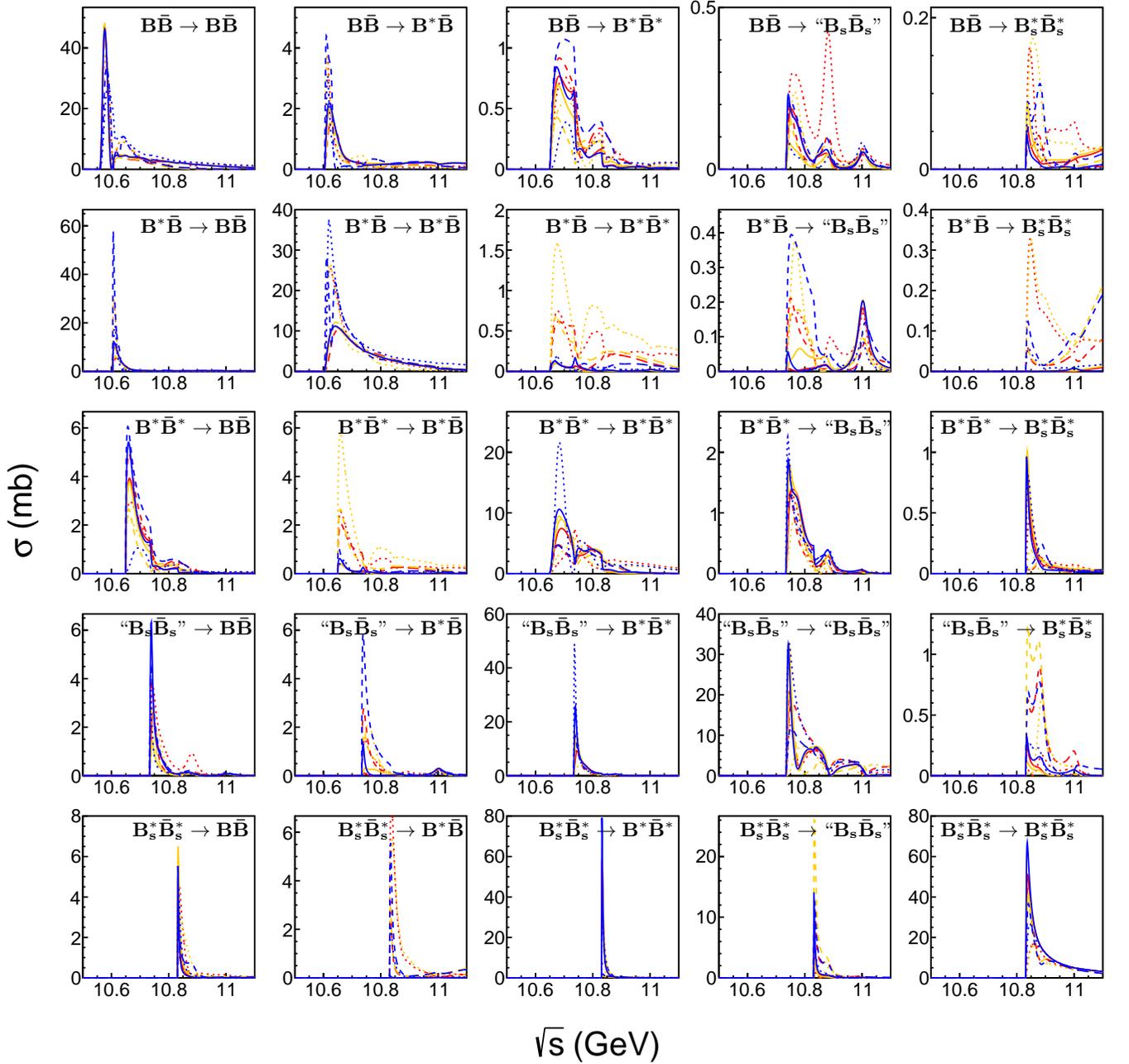
\begin{figure}[h!]
    \centering
    \begin{overpic}[width=1.0\textwidth]{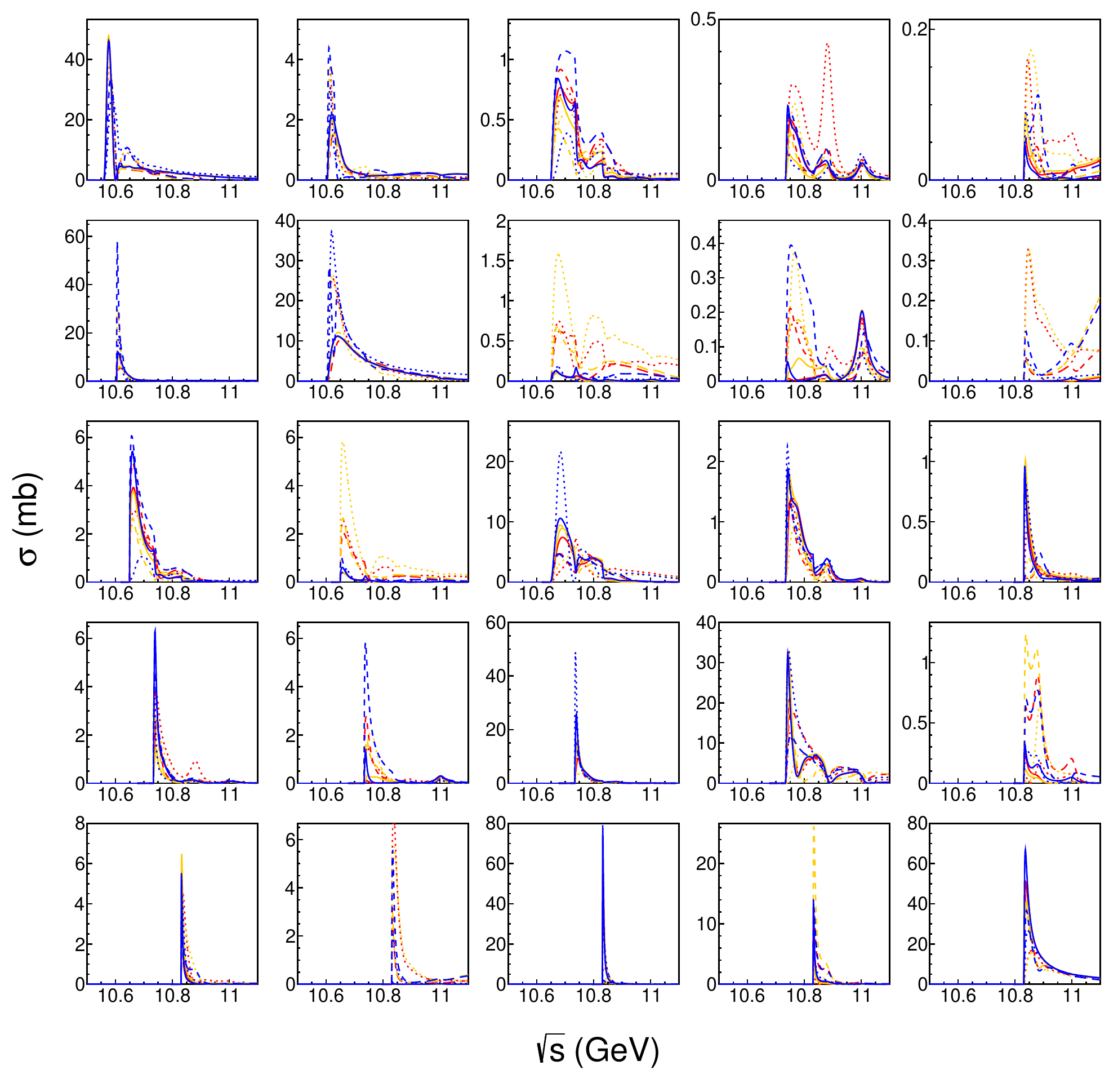}
    \put(13.0,94.5){$\mathbf{B\bar{B}\to B\bar{B}}$}
    \put(13.0,76.5){$\mathbf{B^\ast\bar{B}\to B\bar{B}}$}
    \put(12.5,58.5){$\mathbf{B^\ast\bar{B}^\ast\to B\bar{B}}$}
    \put(11.0,40.5){$\mathbf{``B_s\bar{B}_s"\to B\bar{B}}$}
    \put(12.3,22.4){$\mathbf{B^\ast_s\bar{B}^\ast_s\to B\bar{B}}$}
    
    \put(31.5,94.5){$\mathbf{B\bar{B}\to B^\ast\bar{B}}$}
    \put(31.3,76.5){$\mathbf{B^\ast\bar{B}\to B^\ast\bar{B}}$}
    \put(30.7,58.5){$\mathbf{B^\ast\bar{B}^\ast\to B^\ast\bar{B}}$}
    \put(29.0,40.5){$\mathbf{``B_s\bar{B}_s"\to B^\ast\bar{B}}$}
    \put(30.7,22.4){$\mathbf{B^\ast_s\bar{B}^\ast_s\to B^\ast\bar{B}}$}
    
    \put(49.5,94.5){$\mathbf{B\bar{B}\to B^\ast\bar{B}^\ast}$}
    \put(49.3,76.5){$\mathbf{B^\ast\bar{B}\to B^\ast\bar{B}^\ast}$}
    \put(48.7,58.5){$\mathbf{B^\ast\bar{B}^\ast\to B^\ast\bar{B}^\ast}$}
    \put(47.0,40.5){$\mathbf{``B_s\bar{B}_s"\to B^\ast\bar{B}^\ast}$}
    \put(48.7,22.4){$\mathbf{B^\ast_s\bar{B}^\ast_s\to B^\ast\bar{B}^\ast}$}
    
    \put(67.5,94.5){$\mathbf{B\bar{B}\to ``B_s\bar{B}_s"}$}
    \put(67.3,76.5){$\mathbf{B^\ast\bar{B}\to ``B_s\bar{B}_s"}$}
    \put(66.7,58.5){$\mathbf{B^\ast\bar{B}^\ast\to ``B_s\bar{B}_s"}$}
    \put(65.0,40.5){$\mathbf{``B_s\bar{B}_s"\to ``B_s\bar{B}_s"}$}
    \put(66.7,22.4){$\mathbf{B^\ast_s\bar{B}^\ast_s\to ``B_s\bar{B}_s"}$}
    
    \put(85.5,94.5){$\mathbf{B\bar{B}\to B^\ast_s\bar{B}^\ast_s}$}
    \put(85.3,76.5){$\mathbf{B^\ast\bar{B}\to B^\ast_s\bar{B}^\ast_s}$}
    \put(84.7,58.5){$\mathbf{B^\ast\bar{B}^\ast\to B^\ast_s\bar{B}^\ast_s}$}
    \put(84.7,40.5){$\mathbf{``B_s\bar{B}_s"\to B^\ast_s\bar{B}^\ast_s}$}
    \put(84.7,22.4){$\mathbf{B^\ast_s\bar{B}^\ast_s\to B^\ast_s\bar{B}^\ast_s}$}
    
    \end{overpic}
    \caption{Cross sections for $B_{(s)}^{(\ast)}\bar{B}_{(s)}^{(\ast)} \to  B_{(s)}^{(\ast)}\bar{B}_{(s)}^{(\ast)}$ scattering. Linestyles and colors match those in Fig.~\ref{fig:fitresult2}.
   }
    \label{fig:bbtobb2}
\end{figure}
\end{widetext}

Panel k of Fig. \ref{fig:fitresult2} focuses on the $\Upsilon(4S)$ region in the inclusive cross section. One sees an asymmetric peak and a node near 10.6 GeV that is apparently associated with the $B^*\bar{B}$ threshold (indicated as a dashed line in the panel). We have already noted that the RPP reports an $\Upsilon(4S)$ mass that is approximately 15 MeV below ours. In fact, the real part of our pole position does not lie at the peak of the cross section. The older model, from which the RPP mass is obtained, assumed a Breit-Wigner resonance amplitude (so that the mass necessarily lies near the peak position of approximately 10.58 GeV).
Of course, the assumption of a  Breit-Wigner form can be improved, as we have done with the $K$-matrix formalism. The node appears to be due to destructive interference  between the resonance portion of the amplitude and the background scattering, which turns on just at $B^*\bar{B}$ threshold. In the end, unitarity of the model and this effect combine to yield a pole position that is somewhat removed from the Breit-Wigner mass.

Lastly, Table \ref{tab:chi2} gives the global chi-squared for the nine models considered here. All values lie between 1.08 and 1.19, indicating that the fits are both good, given the disagreement between the inclusive and the sum of exclusive cross sections around 10.65 GeV, and comparable with each other. Again, this shows that model variation is consistent with data variation as revealed in the bootstrap fits, which demonstrates both the reliability of the approach and its limits.

\subsection{Masses}

Pole positions are the most basic quantities extracted from the fit, it is thus of interest to compare our results to quark models and lattice field theory computations. There is a large model literature, which we cannot possibly summarize here. We therefore focus on five models, two recent examples from the literature and three further models that are due to us. These are

\begin{enumerate}
\item (GM) a version of the Godfrey-Isgur ``relativized" quark model, which assumes relativistic kinetic energies,  smeared potentials based on the one gluon exchange interaction, and a Lorentz scalar linear confining potential~\cite{Godfrey:2015dia}.
\item (SOEF) a nonrelativistic constituent model with a frozen coupling, regulated one gluon exchange short range interactions, and a mixture of Lorentz scalar and vector screened confinement potentials~\cite{Segovia:2016xqb}. Use of a screened confinement potential limits the model to low-lying states.
\item (NR) a nonrelativistic model with perturbative spin-dependent interactions reported in Ref.~\cite{Barnes:2005pb}. This model was originally used to describe the charmonium spectrum and was re-fit to 17 well-known bottomonium states here.

\item (ARM) a relativized model reported in Ref.~\cite{Godfrey:2016nwn}. The model uses relativistic kinetic energy and spin-dependent interactions with a Lorentz structure informed by lattice field computations. The spin-dependent interactions are given by (this corrects typographic errors in the original)
\begin{eqnarray}
V_1 &=& -(1-\epsilon) br \nonumber \\
V_2 &=& \epsilon br - C_F \frac{\alpha_S}{r} \nonumber \\
V_3 &=& 3 C_F \frac{\alpha_h}{r^3} \nonumber \\
V_4 &=& C_F \alpha_h \frac{b_h^2}{r} \exp(-b_h r).\nonumber
\end{eqnarray}
The model reported here uses $\epsilon = 0.25$ as obtained by lattice computations and was fit to 59 meson masses across all flavor sectors.

\item (bbg) a simple spin-independent model of mesons and constituent gluon hybrid states, first reported in Ref.~\cite{Farina:2020slb}, where it was applied to charmonia and charmonium hybrids. We have refit the model to 17 bottomonium states here.
\end{enumerate}

The final comparison will be made to the recent lattice field calculation of Ryan and Wilson~\cite{Ryan:2020iog}. This work uses 2+1 dynamical quark flavors on anisotropic lattices of size $20^3\times 128$ and $24^3\times 128$. The pion mass is relatively heavy at  391 MeV. The authors are able to identify the likely composition of states by evaluating  overlaps with a variety of operators. Their preferred identifications are shown in the last column of Table \ref{tab:pdg-qm}.

\begin{widetext}
$\quad$
\begin{table}[ht]
\begin{tabular}{l|lc|lllll|l}
\hline\hline
state & RPP & our estimate & GM & ARM & NR &   bbg  & SOEF & LGT \\
\hline
$1{}^3S_1$ & 9460 &   & 9465 & 9444  & 9454  &  9445 & 9502  & 9419.1(4)\\ 
$2{}^3S_1$ & 10023 & & 10003 & 10029 &  10010   & 10002 & 10015 & 9981(4) \\
$3{}^3S_1$ & 10355 &  & 10354 & 10374 & 10344     & 10339  & 10349  & 10384(12)\\
$4{}^3S_1$ & 10579 & (10590 - 10610) & 10635 & 10641 & 10614    & 10610 & 10607  &   \\
$5{}^3S_1$ & 10885 & (10878 - 10884) & 10878 & 10865 & 10849      & 10848  & 10818  &   \\
$6{}^3S_1$ & 11000 & (11000 - 11008) & 11102 & 11065 & 11064   & 11064 & 10995  &  \\
\hline
$1{}^3D_1$ &      &       &  10138     &  10156     & 10146     &   10148    & 10117 & 10191(9)   \\
$2{}^3D_1$ &      &       &   10441    & 10453      & 10432      &    10435  &  10414 &  10718(33) \\
$3{}^3D_1$ &      &       & 10698      & 10697      & 10679     &    10684    &  10653 &    \\
\hline
$\Upsilon(10750)$ & 10753 & (10630 - 10780) &    &     &      &      & &    \\
\hline
hybrid     &       &       &    &    &      &   11093  &   &  10952(33) \\
\hline\hline
\end{tabular}
\caption{Experimental and Theoretical Vector Bottomonium Masses (MeV).}
\label{tab:pdg-qm}
\end{table}
\end{widetext}

The table reveals that model predictions have typical deviations of tens of MeV from each other, and from experimental masses, with the situation getting worse as one moves up the spectrum. Certainly, it appears that the $\Upsilon(6S)$ is anomalously light (by 60-100 MeV) with respect to theory expectations. The lightest four masses extracted from lattice field theory line up fairly well (within several tens of MeV) with quark models. The authors of Ref.~\cite{Ryan:2020iog} identify the next two vector states as the $(2D)$ and a hybrid meson, as indicated in the table. 

On the experimental side, the authors of Ref.~\cite{MIZUK19} speculate that the $\Upsilon(10750)$ is either a $(3D)$ bottomonium state or a bottomonium hybrid. The approximately 50 MeV deviation from the $(3D)$ quark model is perhaps not unusual; on the other hand, it is  rather far removed from the $(2D)$ state as predicted by the quark models. In contrast, the lattice computation did not resolve a  $D$-wave near 10420 MeV and the authors assigned the 10718 state to $(2D)$, which is a possible attribution for the $\Upsilon(10750)$ if the lattice results are confirmed at lighter quark masses. Finally, the lattice and model calculations yield a vector hybrid near 11000 MeV, which is too heavy for a reasonable interpretation of the $\Upsilon(10750)$ signal.

\subsection{Couplings}
\label{sect:couplings}

Constituent quark model computations of electronic widths and two- and three-body decay modes exist. (In principle these can also be obtained in lattice calculations, but this remains to be done.)  

Table \ref{tab:Gee} shows the electronic partial widths for the resonances considered in this study as reported in the RPP, determined by us, and the quark model computations due to Godfrey and Moats~\cite{Godfrey:2015dia} and Segovia \textit{et al.}~\cite{Segovia:2016xqb}. We also give results from a quark model calculation that attempts to account for persistent over-estimation of decay constants in nonrelativistic models by softening the short range interaction with the aid of the running coupling~\cite{Lakhina:2006vg} (labelled ``LS" in the table).

\begin{table}[ht]
\begin{tabular}{l|lc|lll}
\hline\hline
state & RPP & our estimate & LS & GM & SOEF \\
\hline
$\Upsilon(4S)$ & $0.272$ & (0.003 - 0.62) & 0.31 & 0.39 & 0.21 \\
$\Upsilon(5S)$ & $0.31$ & (0.037 - 0.068) & 0.28 & 0.33 & 0.18 \\
$\Upsilon(6S)$ & $0.13$ & (0.043 - 0.074) & 0.26 & 0.27 & 0.15 \\
$\Upsilon(10750)$ & (0.01 - 0.40)\footnote{from ambiguous solutions in Ref.~\cite{DONG20A}} & (0.004 - 0.10) &  & 2.38 eV \footnote{assuming a $3D$ state} &  \\
\hline\hline
\end{tabular}
\caption{Experimental and Theoretical $e^+e^-$ Widths (keV). The  entries in ``our estimate" correspond to the full range of model results.} 
\label{tab:Gee}
\end{table}

The table suggests that the quark models tend to agree on the magnitude of the coupling and that, perhaps not surprisingly, these roughly agree with the measured values as reported in the RPP. However, as discussed above, typical older model assumptions lead to widths 
that are higher than they should be. Indeed, our results are smaller by approximately a factor of two for the $\Upsilon(6S)$ and a factor of six for the $\Upsilon(5S)$. The value for $\Upsilon(10750)$ is new. The trend is that these widths (and hence, decay constants) fall more rapidly with radial quantum number than previously thought, which implies that short range dynamics in typical constituent quark models is not as well understood as hoped. Future lattice field theory computations of decay constants should help clarify the situation.

We extract an $e^+e^-$ partial width of (4 - 100) eV for the $\Upsilon(10750)$. For most models and bootstrap solutions, we find values substantially larger than the roughly 2 eV that quark models typically obtain for $D$-wave widths. Possible explanations are that (i) $S$-$D$ wave mixing is larger than typical quark models predict, (ii) the $D$-wave identification is incorrect, or (iii) quark model descriptions of short range dynamics become worse as the angular momentum increases.
In view of the large overall spread in our fit results, it is perhaps too early to speculate on this discrepancy; although we suspect that option (iii) should be considered further by modellers.

Tables \ref{tab:4SBR} - \ref{tab:6SBR} report branching fractions for the four resonances as determined here, RPP values, and theoretical expectations from the GM and SOEF quark models.

As expected, in most of the models and bootstraps, nearly the entire width of the $\Upsilon(4S)$ is in the $B\bar{B}$ mode (Tab. \ref{tab:4SBR}). As discussed above, we do however report a nonzero value for the $B^*\bar{B}$ mode because a number of bootstrap poles lie above this threshold. Our three-body branching fractions are two to four times larger than those given in the RPP; however our ranges are quite large.
These higher branching fractions are a consequence of the pole position being shifted towards higher masses. The three-body branching fractions are constrained by a single Belle data-point 
on the $\Upsilon(4S)$ resonance. If the pole is indeed shifted, the peak in the cross sections of the three-body processes is shifted as well, leading to a larger branching ratio when compared with the assumption, that the energy point is in the maximum of the resonance.
The quark model three-body branching fractions are  in the ranges of both the RPP and our measurements, and it is perhaps too early to make judgements on the efficacy of the models.

\begin{table}[ht]
\begin{tabular}{l|ll|ll}
\hline\hline 
channel & RPP & our estimate & GM   & SOEF \\
\hline
$B\bar{B}$ & $>$ 96 & (66 - 100) & $\approx $ 100 &
$\approx$ 100 \\
$B^*\bar{B}$ &  & (0.02 - 33) & 0 & 0 \\
$\Upsilon(1S)\pi\pi$ & $8.2\cdot 10^{-3}$ & (0.0001 - 0.032) & $7.5\cdot 10^{-3}$ & $29.4\cdot 10^{-3}$ \\
$\Upsilon(2S)\pi\pi$ & $8.2\cdot 10^{-3}$ & (0.002 - 0.048) & $8.0\cdot 10^{-3}$ & $1.2 \cdot 10^{-3}$\\
\hline\hline
\end{tabular}
\caption{$\Upsilon(4S)$ Branching Ratios (\%). The entries in ``our estimate" correspond to the full range of model results.}
\label{tab:4SBR}
\end{table}

Extracted branching ratios and quark model predictions for the $\Upsilon(10750)$ are presented in Tab. \ref{tab:750BR}, including model predictions for the $(3D)$ state.
Within the large uncertainties, there is a reasonably good match with the quark model predictions, lending some support to the identification of the $\Upsilon(10750)$ with the $\Upsilon(3D)$. 


\begin{table}[ht]
\begin{tabular}{l|ll|ll}
\hline\hline
channel & RPP & our estimate & GM  & SOEF \\
\hline
$B\bar{B}$     &  -- & (0.7 - 58) & 23  & --    \\
$B^*\bar{B}$   &  -- & (0.1 - 67) & 0.2 &  --   \\
$B^*\bar{B}^*$ &  -- & (0.7 - 59) & 76.7 &  -- \\
``$B_s\bar{B}_s$" & --  & (0.008 - 89)  & --  & --  \\
$B_s^*\bar{B}_s^*$ & --  & (0.05 - 34) & --  & --  \\
$\Upsilon(1S)\pi\pi$ &--  & (0.005 - 1.8) &-- & -- \\
$\Upsilon(2S)\pi\pi$ & -- & (0.009 - 3.8) & -- & -- \\
$\Upsilon(3S)\pi\pi$ &--  & (0 - 0.6) & -- &  \\
$h_b(1P)\pi\pi$ & --  & (0.01 - 2.6) &-- & -- \\
$h_b(2P)\pi\pi$ & -- & (0.004 - 1.7) & -- & -- \\
\hline\hline
\end{tabular}
\caption{$\Upsilon(10750)$ Branching Ratios (\%).}
\label{tab:750BR}
\end{table}

\begin{table}[ht]
\begin{tabular}{l|ll|ll}
\hline\hline
channel & RPP\footnote{obtained using ratios of cross sections near the $\Upsilon(5S)$ mass} & our estimate & GM    & SOEF \\
\hline
$B\bar{B}$ & 5.5   & (0.6 - 31) & 19.5 & 22.3   \\
$B^*\bar{B}$   & 14  & (0.03 - 3.2) & 60.6 &  42.4  \\
$B^*\bar{B}^*$ & 38  &  (2.9 - 17) & 8.8 & 0.3     \\
``$B_s\bar{B}_s$" & [25]  & (31 - 77) & --  & --  \\
$B_s^*\bar{B}_s^*$ & 18  &  (0.9 - 33)  & 7.3  & 27.4  \\
$\Upsilon(1S)\pi\pi$ & 0.53 & (0.6 - 2.5) & -- & 0.023 \\
$\Upsilon(2S)\pi\pi$ & 0.78 & (1.6 - 5.2) & -- & 0.033 \\
$\Upsilon(3S)\pi\pi$ & 0.48 & (0.2 - 1.7) & -- & 0.01 \\
$h_b(1P)\pi\pi$ & 0.35 & (0.3 - 2.8) & -- & -- \\
$h_b(2P)\pi\pi$ & 0.57 &  (0.9 - 4.0) & -- & -- \\
\hline\hline
\end{tabular}
\caption{$\Upsilon(5S)$ Branching Ratios (\%).}
\label{tab:5SBR}
\end{table}

Table \ref{tab:5SBR} shows our results for the $\Upsilon(5S)$. We see that the missing channel determined here is up to three times larger than that reported in the RPP. We have attributed this substantial missing strength to $B^*\bar{B}^{(*)}\pi$, potentially through the $Z_b^{(\prime)}$ intermediate state; although we remind the reader that experiments reported ratios of cross sections rather than branching fractions in this case.

Other branching fractions also show substantial disagreement with the RPP values, with the largest discrepancy in the $B^*\bar{B}$ channel. Comparison to the quark models is somewhat confused because they do not account for all possible channels in determining the total width. Of course, relative branching fractions remain meaningful. Both models predict that $B^*\bar{B}$ is the dominant decay mode with $B^*\bar{B}^*$ an order of magnitude smaller. 
We find some evidence for the reverse situation, which agrees with the trend in the RPP.

Lastly, we consider the $\Upsilon(6S)$ branching fractions shown in Tab. \ref{tab:6SBR}. Once again, much of the full width appears to be unmeasured, with approximately 80\% going to the missing channel. We attribute this to missing $B^*\bar{B}^{(*)}\pi$, as discussed above. In this case, the quark models agree on the ordering $Bf(B^*\bar{B}) \gtrsim Bf(B^*\bar{B}^*) > Bf(B\bar{B})$. We, on the other hand, find $Bf(B^*\bar{B}) \gg Bf(B^*\bar{B}^*)$, although we stress that new data on exclusive cross sections above 11 GeV could lead to significant changes in these branching fractions.

Although branching fractions are not reported for the $\Upsilon(10750)$ and $\Upsilon(6S)$ in the RPP, they list ratios of widths, as shown in Tables \ref{tab:750-ratios} and  \ref{tab:6S-ratios}. Our results for the $\Upsilon(6S)$ align very well with those reported in the RPP, which supports our estimated branching fractions in Tab. \ref{tab:6SBR}. Alternatively, our ratio for the $\Upsilon(3S)\pi\pi$ decay mode of the $\Upsilon(10750)$ is smaller than that of the RPP over the full model range. The three-body branching fractions of Tables \ref{tab:750BR}, \ref{tab:5SBR}, and \ref{tab:6SBR} are all comparable in size, which we find reasonable.

Finally, we draw attention to the large branching fractions for closed bottom three-body channels for the $\Upsilon(5S)$ (Tab. \ref{tab:5SBR}) and $\Upsilon(6S)$ (Tab. \ref{tab:6SBR}), which are three or more orders of magnitude larger than those seen for the $\Upsilon(4S)$. The formerly perplexing dipion decays of heavy quarkonia are now widely believed to be understood in terms of long-lasting open-bottom fluctuations and mixing with the novel $Z_b$ states (see, for example, Ref.~\cite{Chen:2019gty}). The challenge here  is that the $Z_b$ mesons have masses that are rather far removed from the $\Upsilon(5S)$ and $\Upsilon(6S)$ masses. Relevant box diagrams contain $\bar{B}^{(*)} B^{(*)}$ mesons and are therefore also far removed in mass. Of course the possibility of higher mass resonances in the box diagram is open, and should be investigated.

We are aware of one computation that uses an effective field theory approach to estimate dipion transition rates for $\Upsilon(10750)$ and $\Upsilon(6S)$ assuming that these states are hybrids~\cite{TarrusCastella:2021pld}. Predictions are $\Gamma(\Upsilon(10750) \to \Upsilon(1S)\pi\pi) \approx 43.4$ keV  and $\Gamma(\Upsilon(6S) \to \Upsilon(1S)\pi\pi) \approx 99.1$ keV, with substantially smaller widths for $\Upsilon(2S,3S)\pi\pi$. Evidently our branching fractions deviate significantly from these predictions, leading one to suspect either the hybrid designation of these states or the accuracy of the calculation assumptions.

In summary, it appears that the model predictions for  open-bottom decay 
widths for $\Upsilon(5S)$ and $\Upsilon(6S)$ do not agree well with our extracted residues, although the uncertainties are large. Both groups (GM~\cite{Godfrey:2015dia} and SOEF~\cite{Segovia:2016xqb})
used the well-established ``3p0" model for strong decays to obtain their results. Although the model is very simple, the ratios of partial widths that are related by spin symmetry should be  robust. The lack of agreement could then imply that the assumed spin structure of the 3p0 model is incorrect. An alternative is that these partial widths also depend on wavefunction overlaps, and these can be particularly difficult to model high in the spectrum, where details such as the location of wavefunction nodes can become very important.

\begin{table}[ht]
\begin{tabular}{c|ll|ll}
\hline\hline
channel & RPP & our estimate & GM    & SOEF \\
\hline
$B\bar{B}$     &  -- & (0.8 - 8.6) & 3.9 & 5.3   \\
$B^*\bar{B}$   &  -- & (1.9 - 12) & 22.4 & 19.6    \\
$B^*\bar{B}^*$ &  -- & (0.2 - 6.2) & 17.4 & 15.0  \\
``$B_s\bar{B}_s$" & --  & (70 - 90)  & --  & --  \\
$B_s^*\bar{B}_s^*$ & --  & (0.04 - 9.7) & 0.9  & 2.6  \\
$\Upsilon(1S)\pi\pi$ &--  & (0.3 - 1.2) & -- & 0.35 \\
$\Upsilon(2S)\pi\pi$ & -- & (0.3 - 2.9) & -- & $8.0 \cdot 10^{-3}$ \\
$\Upsilon(3S)\pi\pi$ &--  & (0.2 - 1.0) & -- & 0.049 \\
$h_b(1P)\pi\pi$ & --  & (0.5 - 2.1) & -- & -- \\
$h_b(2P)\pi\pi$ & -- & (0.2 - 4.3) & -- & -- \\
\hline\hline
\end{tabular}
\caption{$\Upsilon(6S)$ Branching Ratios (\%).}
\label{tab:6SBR}
\end{table}

\begin{table}[ht]
\begin{tabular}{l|lc}
\hline\hline
channel & RPP & our estimate \\
\hline
1S & 0.295(175) & (0.0003 - 0.74) \\
2S & 0.875(275) & (0.001 - 2.7) \\
3S & 0.235(25)  & (0 - 0.16) \\
\hline\hline
\end{tabular}
\caption{$\Gamma(\Upsilon(10750)\to e^+e^-) \cdot \Gamma(\Upsilon(10750) \to \Upsilon(nS) \pi\pi)/\Gamma_{\mathrm tot}$ (eV).}
\label{tab:750-ratios}
\end{table}

\begin{table}[ht]
\begin{tabular}{l|lc}
\hline\hline
channel & RPP & our estimate \\
\hline
1S & 0.46(8)  & (0.19 - 0.71) \\
2S & 0.65(52) & (0.13 - 2.0) \\
3S & 0.33(16) & (0.12 - 0.69) \\
\hline\hline
\end{tabular}
\caption{$\Gamma(\Upsilon(6S)\to e^+e^-) \cdot \Gamma(\Upsilon(6S) \to \Upsilon(nS) \pi\pi)/\Gamma_{\mathrm tot}$ (eV).}
\label{tab:6S-ratios}
\end{table}

\section{Conclusions}
\label{sect:concl}

We have provided the first comprehensive analysis of vector bottomonia using data on both the production of various exclusive open- and hidden-bottom channels and the inclusive cross section for bottom anti-bottom quark pair production in electron positron annihilation. In a unitary approach using the $K$-matrix formalism, we find a total of four poles, the $\Upsilon(4S)$, $\Upsilon(10750)$, $\Upsilon(5S)$ and $\Upsilon(6S)$ in accordance with the RPP. Allowing for a non-resonant contribution and a proper treatment of thresholds, we extract a pole position of the $\Upsilon(4S)$ that is about 10-20 MeV higher than previous values. While we find a significant contribution of the additional $\Upsilon(10750)$ state previously observed by Belle in $e^+e^-\to \Upsilon(nS)\pi^+\pi^-$, the paucity of the data in that energy region allows for a broad range of masses and widths, all describing the data similarly well. The large number of two-body thresholds in the $\Upsilon(4S)$ and $\Upsilon(10750)$ regions further complicate the situation. The pole positions of the $\Upsilon(5S)$ and $\Upsilon(6S)$ are well constrained by the (inclusive) data, with the $\Upsilon(5S)$ position agreeing well with the RPP average and the $\Upsilon(6S)$ being about twice as wide as previous measurements.

The $K$-matrix description allows us to determine absolute branching ratios of all four states for the first time, although relatively few data points in the exclusive measurements and (potentially as a consequence) a significant model dependence lead to large uncertainties. However, our method of extracting these branching fractions is more robust than the ratios of cross sections currently being reported in the RPP, especially given that none of the exclusive cross sections of open-bottom production exhibit clean peaks in the $\Upsilon(10750)$, $\Upsilon(5S)$ and $\Upsilon(6S)$ regions. For these three states, we find a large fraction of the intensity is still missing, with $B^{(*)}\bar{B}^{(*)}\pi$ channels, potentially populated by $Z_b^{(\prime)}$ contributions, identified as promising candidates to explain the discrepancy between the exclusive and the inclusive data. A measurement of these cross sections using Belle(II) data could resolve this issue in the future, thereby significantly reducing model uncertainties.

Additional data in the $\Upsilon(10750)$ region from Belle~II is highly anticipated in order to more cleanly study this state, with a more precise measurement of its branching ratios aiding in identifying its nature. Similarly, additional exclusive measurements in the $\Upsilon(6S)$ region above 11 GeV would be beneficial to reduce current ambiguities caused by the total absence of such data.
Finally, one might speculate about the nature of the $\Upsilon(5S)$ and $\Upsilon(6S)$ states, given that their decay rate to three-body hidden-bottom channels is significantly enhanced compared to the conventional $\Upsilon(4S)$ state. Whether this can be explained by a strong coupling to the exotic $Z_b^{(\prime)}$ states or requires an exotic interpretation of the $\Upsilon(5S)$ and $\Upsilon(6S)$ states remains to be answered.

Comparison with quark model calculations reveal that bottomonium masses are in reasonable agreement with our pole positions. However, there are indications that partial widths to open bottom channels are in disagreement with experiment, implying that the decay model may need to be revisited or that details in the computation are incorrect. Similarly, our electronic widths deviate substantially from RPP values and model predictions, showing a rapid fall with radial quantum number. This observation is difficult to explain in simple quark models, and hint at the necessity of radical modification to models if the results are confirmed in the future.

\section{Acknowledgements}

NH acknowledges support from the European Union Horizon 2020 research and innovation programme under Marie Sk\l{}odowska-Curie grant agreement No 894790. NH wishes to thank M. Albrecht and B. Kopf for many helpful discussions on the $K$-matrix formalism. Swanson's research was supported by the U.S. Department of Energy under contract DE-SC0019232.  Mitchell's research was supported by the U. S. Department of Energy under Contract DE-FG02-05ER41374.

\appendix

\section{K-matrix Formalism}
\label{appA}

A perusal of the literature reveals a plethora of conventions in use for the $K$-matrix formalism, which can lead to confusion. For this reason we present a somewhat pedagogic review of the method with explicit formulas for a variety of relevant quantities. A critique of some common model choices (along with a description of our modeling) is also presented.

We start with the conventional relationship between the scattering matrix ($S$) and the transition matrix ($T$):

\be
S = 1 + iT.
\ee
Conservation of probability is equivalent to unitarity of the $S$-matrix  (namely $SS^\dagger = S^\dagger S = 1$), which implies that
\be
T -  T^\dagger = i T T^\dagger = i T^\dagger T.
\label{eq:Toptical}
\ee
To facilitate using the conventional scattering formulas of quantum field theory (as exemplified, for example in the RPP), we choose to work with the invariant amplitude~($\mathcal{M}$) defined by

\be
T = (2\pi)^4 \delta^4(p_i^{tot} - p_f^{tot}) \prod_{f} \frac{1}{\sqrt{2 E_{f}}} \cdot \mathcal{M} \cdot \prod_i \frac{1}{\sqrt{2 E_{i}}},
\ee
where the index $f(i)$ refers to the final (initial) particles in the reaction.

In this case Eq. \ref{eq:Toptical} becomes 
\begin{eqnarray}
-i(\M_{\mu\nu} - \M^*_{\nu\mu}) &=& \sum_\gamma \frac{1}{S_\gamma} \int \M_{\mu\gamma}(\{p_\mu\}, \{k_\gamma\}) \\ &\cdot& \M^*_{\nu\gamma}(\{p_\nu\},\{k_\gamma\}) \, d\Phi(\{p_\mu\}:\{k_\gamma\}), \nonumber
\label{eq:MM}
\end{eqnarray}
where we now make the channel indices explicit and introduce a symmetrization factor, $S_\gamma$. The right hand side comes from $\sum_\gamma T_{\mu\gamma}T^*_{\nu\gamma}$ where the notation is resolved as a sum over channels (denoted $\gamma$) and an integral over three-momentum. 
The result is written in terms of  the invariant phase space $d\Phi$ (this is $(2\pi)^4$ larger than the RPP convention) defined by
\be
d\Phi(\{p_i\}:\{k_f\}) \equiv (2\pi)^4 \delta^4(p_i^{tot}- k_f^{tot}) \prod_f \frac{d^3 k_f}{(2\pi)^3 2 E_f}.
\ee

For two bodies $A\to C+D$ in the center of mass frame 
\begin{eqnarray}
d\Phi &=& (2\pi)^4 \delta^4(p_C+p_D - p_A) \frac{d^3 p_C}{(2\pi)^3 2 E_C} \frac{d^3 p_D}{(2\pi)^3 2 E_D}\nonumber \\ &=& \frac{1}{4\pi^2} \frac{k_*}{4 M_A} d\Omega.
\end{eqnarray}
We have introduced the center of mass momentum $k_*(M_A^2)$ which we generalize to  

\be
k_*(s) = \frac{\sqrt{(s-(m_1+m_2)^2)(s-(m_1-m_2)^2)}}{2\sqrt{s}}.
\ee
We stress that this quantity will not be considered below threshold in the following (except when exploring the $T$-matrix in the complex plane) for reasons to be discussed shortly. Thus it is implemented with a theta function forcing it to zero below threshold. 

In keeping with particle physics convention, the branch cut in the square root will be taken along the positive real axis (however, the usual branch cut along the negative real axis will be used for the square root in the denominator, which exhibits a non-physical singularity at $s=0$).


Now restrict attention to two-body scattering in the center of mass frame with two-body intermediate states $\gamma$. We get
\begin{eqnarray}
\M_{\mu\nu}-\M^*_{\nu\mu} &=& i\sum_{\gamma} \frac{1}{S_\gamma} \int \M_{\mu\gamma}(\{p_\mu\}, \{k_\gamma\}) \\ &\cdot& \M^*_{\nu\gamma}(\{p_\nu\},\{k_\gamma\}) \,  \frac{k_*^\gamma}{16 \pi^2 \sqrt{s}} d\Omega(k_1). \nonumber
\end{eqnarray}
Finally, assume that the amplitudes are not functions of angle so that this equation reduces to an algebraic equation. Going to matrix form then gives:
\be
\M - \M^\dagger = 2 i \M \rho \M^\dagger
\label{eq:MM2}
\ee
where
\be
(\rho)_{\mu\gamma} \equiv \delta_{\mu\gamma}  \frac{k_*^\gamma}{S_\gamma 8 \pi \sqrt{s}} 
\label{eq:rho}
\ee
is the phase space matrix. Different conventions for the phase space exist -- these can be absorbed into the definition of $K$ and will not affect pole positions but will yield unconventional expressions for cross sections and other physical quantities. 

Eq. \ref{eq:MM2} is equivalent to 
\be
(\M^{-1}+ i \rho)^\dagger = \M^{-1} + i \rho,
\ee
which implies that the quantity in brackets is real and symmetric. It is therefore useful to identify a matrix with the same properties:

\be
K^{-1} \equiv \M^{-1} + i \rho - R,
\ee
where $R$ is a real function along the diagonal of $K$.  The function $R$ can in turn be incorporated into the phase space by defining a new quantity, $C$, as
\be
C \equiv R-i\rho.
\ee
Thus we have
\be
\M^{-1} = K^{-1} + C
\ee
or
\begin{eqnarray}
\M &=& (1+KC)^{-1}K = K(1+CK)^{-1}\nonumber \\ &=& K(K+KCK)^{-1}K.
\label{eq:M}
\end{eqnarray}

Many choices for the function $C$ are possible (and many are made in the literature). However it is preferable to use one that maintains analyticity across thresholds (hence $R \ne 0$) since this smooths discontinuities that fitting routines find difficult to handle. 
In this paper we will therefore set $C$ equal to the Chew-Mandelstam function~\cite{Wilson:2014cna,albrecht}. The real part of the Chew-Mandelstam function is related to its imaginary part by a once-subtracted dispersion integral,
\be
C_\gamma(s) = C(s_\gamma) - \frac{s-s_\gamma}{\pi} \int_{s_\gamma}^\infty ds'\, \frac{r(s')}{(s'-s)(s'-s_\gamma)},
\ee
that ensures a smooth transition across threshold. Here $s_\gamma = (m_1+m_2)^2$ is the threshold for channel $\gamma$. 
The auxiliary function is chosen to guarantee that the imaginary part of $C_\gamma$ is $\rho_\gamma$. Thus $r$ is the continuation of $\rho$ to the complex plane:

\be
r_\gamma(s) \equiv \frac{ \sqrt{\left(1-\frac{(m_1+m_2)^2}{s+i\epsilon}\right)\cdot\left(1-\frac{(m_1-m_2)^2}{s+i \epsilon}\right)}}{16\pi S_\gamma}
\label{eq:defrho}
\ee

The integral can be done yielding (we set $C(s_\gamma)=0$)~\cite{Wilson:2014cna},

\begin{equation}
C_\gamma(s) = \frac{r_\gamma(s)}{\pi}\log\left[\frac{\xi_\gamma(s)+r_\gamma(s)}{\xi_\gamma(s)-r_\gamma(s)}\right] - \frac{\xi_\gamma(s)}{\pi}\frac{m_2-m_1}{m_2+m_1}\log\frac{m_2}{m_1},
\end{equation}
where the additional function is defined by

\be
\xi_\gamma(s) = \frac{1}{16\pi S_\gamma} \cdot \left(1-\frac{(m_1+m_2)^2}{s+i\epsilon}\right). 
\ee

\begin{figure}[ht]
\includegraphics[width=0.5\textwidth,angle=0]{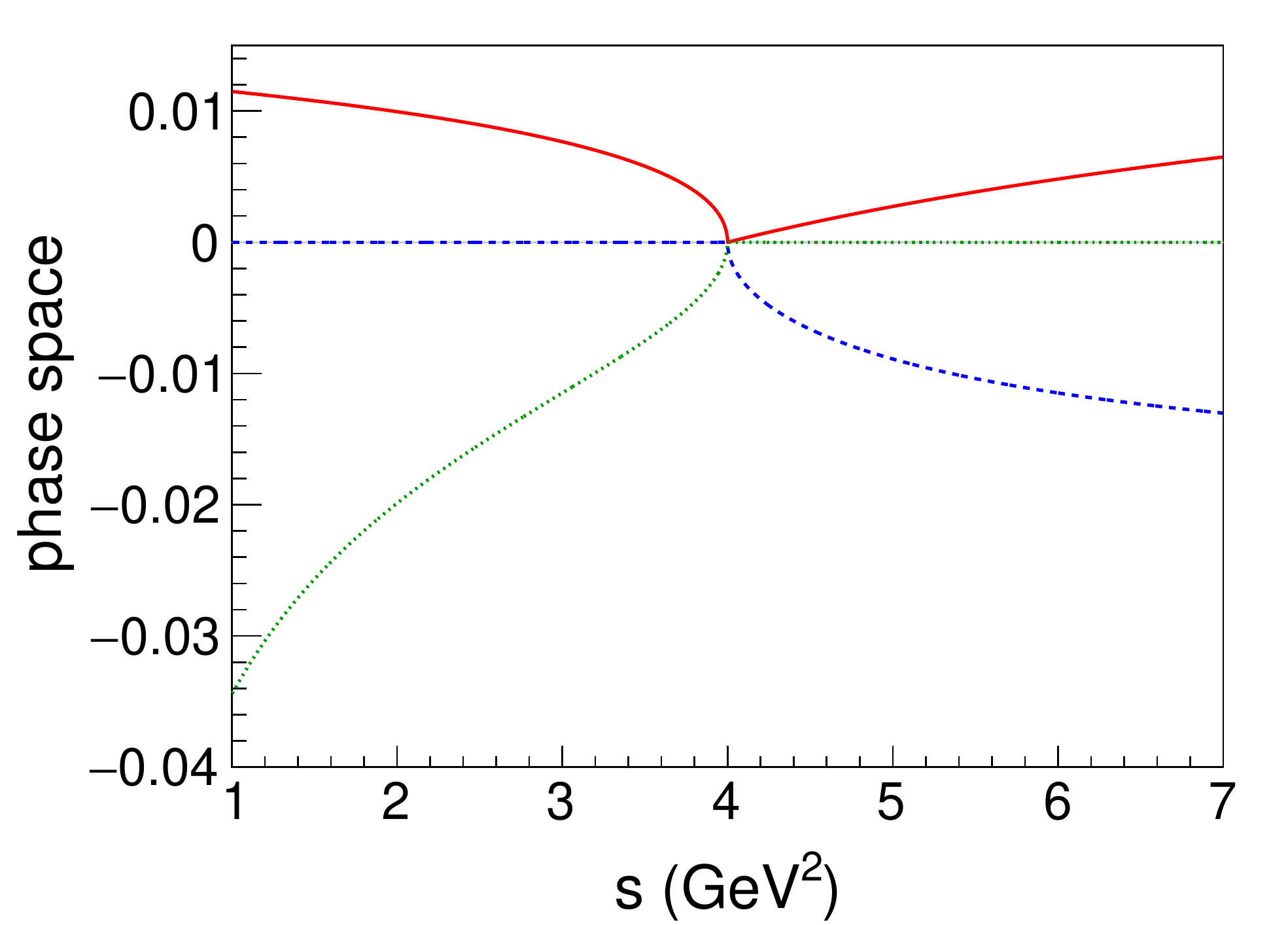}
\caption{The Chew-Mandelstam function for ($m_1=m_2=1$). red solid line: real part of the Chew-Mandelstam function. Blue dashed line: imaginary part of the Chew-Mandelstam function. Green dotted line: imaginary part of $\rho(s)$ continued below threshold.}
\label{fig:CM}
\end{figure}

Fig. \ref{fig:CM} displays the real and imaginary parts of the Chew-Mandelstam function. We remark that 
many authors choose to continue $k_*$ (and thus $\rho(s)$) below threshold in the name of ``analyticity". This amounts to making a model choice of the real part of $C$ that is particularly poor, as illustrated in the figure, and we recommend against this practice.

Some authors choose to include angular momentum factors in the Chew-Mandelstam function~\cite{rpp50}.
The derivation we have presented leaves no room for modifications of this sort. And in fact, it is clear that these factors are associated with the scattering amplitude rather than the phase space, and are better modeled in the couplings, as described below. 

The remaining task is to specify the elements of $K$. 
Gribov has argued that scattering amplitudes factorize near poles~\cite{Gribov:2009zz}, hence it is sensible to write the elements of $K$ in terms of the product of channel couplings and a ``bare" pole. This is common practice in the field, which we follow here. Thus we parameterize $K$ with resonant and non-resonant (``background") terms:

\be
K_{\mu,\nu} = \sum_R \frac{g_{R:\mu} g_{R:\nu}}{m_R^2 -s} + f_{\mu,\nu}. \label{eq:HMSK}
\ee
The index $R$ refers to a resonance and Greek indices refer to continuum channels. The couplings must be real but can be $s$-dependent (angular dependence is excluded by the assumptions leading to Eq. \ref{eq:M}).
Extracted couplings are spin-averaged for $g_{R:ee}$ and are spin-summed in other cases. 

Many choices for the $s$-dependence of the couplings exist. To facilitate the discussion and to gain insight into these choices  it is helpful to compare to the propagator for a scalar field. To this effect we consider a heavy field $\phi$ with mass $M$ coupled to a light scalar with mass $m$. The interaction lagrangian  is taken to be $\frac{1}{2} g \int d^4 x\, \phi \varphi^2$. Performing the standard Dyson sum yields a scattering amplitude of 

\be
\M = \frac{-g^2}{s-M^2-\Pi(s)},
\label{eq:P}
\ee
where $-i\Pi$ is the $\phi$ self-energy given by (we use the $\overline{MS}$ scheme to renormalize):

\be
\Pi(s) = \frac{g^2}{16 \pi^2 S} \, \int_0^1 dx\,  
\log[x m^2 + (1-x) m^2 - x(1-x) s - i \epsilon].
\ee

\noindent
The imaginary part of the self-energy is given by

\be
\Im(\Pi) =  - \frac{1}{16\pi S} g^2\, \sqrt{1 - s_\gamma/s} \cdot \theta(s - s_{\mathrm{thr},\gamma}) = -g^2 \rho_\gamma(s). 
\label{eq:ImPi}
\ee
At lowest order, the $\phi$ particle width is given by

\be
\Gamma = \frac{g^2}{8\pi}\frac{k_*(M)}{S M^2},
\label{pdgG}
\ee
which can be immediately generalized to the case of two-to-two scattering in the center of mass frame as

\be
\Gamma(s) = \frac{g^2 k_*(s)}{8\pi S s}.
\ee
from which we conclude

\be
\Im(\Pi) = -\sqrt{s}\Gamma(s).
\label{eq:G}
\ee

Notice that, because the imaginary part of the self-energy is (up to $-g^2$) the phase space, $\rho$, and because the loop integral is analytic, the Chew-Mandelstam function is related to the self-energy by $C(s) = g^2\Pi(s)$ (up to a constant). One can therefore think of employing the Chew-Mandelstam function as a prescription for bringing the $K$-matrix scattering amplitude into agreement with the Dyson form of the full scalar propagator.

Eq. \ref{eq:G} shows the relationship between the $s$-dependent resonance width and phase space. In general this relationship can be more complicated due to vertex corrections, or, less formally, due to the finite size of the hadrons involved~\cite{B}.
At minimum it is known that $\Gamma(M) \sim k_*(M)^{2\ell+1}$ where $\ell$ is the angular momentum in the final state. This can be incorporated in the formalism by setting $g(s) = g_0 k_*(s)^\ell$. It is of course possible to go further by specifying additional $s$-dependence that is meant to mimic hadronic interactions at the vertex.  Well known examples are the Walker model~\cite{Walker}

\begin{equation}
\Gamma_W(s) = \Gamma_R \left(\frac{k_*}{ k_R}\right)^{2\ell+1} \frac{M}{\sqrt{s}} 
\left(\frac{k_R^2+\beta^2}{ k_*^2+\beta^2}\right)^\ell
\end{equation}
and the similar MAID model~\cite{maid}

\begin{equation}
\Gamma_M(s) = \Gamma_R \left(\frac{k_*}{k_R}\right)^{2\ell+1} \frac{M^2}{s}
\left(\frac{k_R^2+\beta^2}{k_*^2+\beta^2}\right)^\ell.
\end{equation}
The model of von Hipple and Quigg~\cite{vHQ}  incorporates an estimate of  centrifugal barrier penetration and reads

\begin{equation}
\Gamma_{vHQ}(s) = \Gamma_R \frac{k_R}{k_*} \frac{M}{\sqrt{s}} \frac{|h_\ell(k_R/\beta)|^2}{|h_\ell(k_*/\beta)|^2},
\end{equation}
where $h_\ell$ is a spherical Hankel function. This is closely related to the commonly employed Blatt-Weisskopf centrifugal  
barrier penetration factor, which is derived under the assumption that the
semiclassical impact parameter $\sqrt{\ell(\ell+1)}/k$
is much larger than the range of the final state interaction potential, $1/\beta$.
The effects of barrier penetration may then be computed in nonrelativistic quantum 
mechanics by matching the outgoing wavefunction in the $\ell$th wave to an 
assumed inner wavefunction at the distance $1/\beta$.  The resulting transmission
coefficients are called Blatt-Weisskopf factors~\cite{BlW} and are denoted $v_\ell(k/\beta)$.
We note that, as required,  $v_\ell \sim k^{2\ell}$ for small momentum and $v_\ell \to 1$ for large 
momentum (we take $v_0(x) = 1$, $v_1(x) = x^2/(1+x^2)$, 
$v_2(x) = x^4/(9 + 3 x^2 + x^4)$, etc.). The resulting width model can be written as

\begin{equation}
\Gamma_{BlW}(s) = \Gamma_R \frac{k_*}{k_R} \frac{M^2}{s} \frac{v_\ell(k_*/\beta)}{v_\ell(k_R/\beta)}.
\label{eq:blw}
\end{equation}

The general utility of the Blatt-Weisskopf factors may be questioned. For example, the factorization $k_* \ll \beta$ required to enable wavefunction matching
is only true  very near threshold and does not hold over typical energy ranges
involved in data analysis. Indeed the assumption that the strong force is weak
beyond 1 fm is often vitiated by one-pion-exchange forces. Furthermore, the Blatt-Weisskopf and von Hippel-Quigg factors do not
impose any damping on the form factor for $k_* \gg \Lambda_{QCD}$, which is not consistent with expectations for hadronic interactions. Finally, all the models incorporate normalization on resonance. This is convenient for interpreting multiplicative factors (in this case, $\Gamma_R$), however, it is problematic below threshold where $k_R$ does not exist. We will therefore not use this form of normalization in our model.

An alternative way to obtain width form factors is to employ a microscopic model of hadronic decays. For example the ``3p0" strong decay model has been used to obtain a form factor describing $\Upsilon(4S) \to B\bar B$ decay by the BaBar collaboration~\cite{BaBar:2004rrm}. This form factor plays an important role in the analysis as it has a node just beyond the $\Upsilon(4S)$ peak.

Motivated by the same decay model, we choose to parameterize the coupling constants for resonance $R$ and channel $\alpha$ as 

\be
g_{R:\mu}(s) = \hat g_{R:\mu} \left(\frac{k_\mu(s)}{\beta}\right)^{\ell_\mu} \cdot \exp[- k^2_\mu(s)/\beta^2].
\label{eq:gFormFactor}
\ee


Recall that $k_\mu$ is zero below threshold. We choose not to normalize at the point $s=M_R^2$ because the relevant momentum does not exist when the bare resonance is below threshold.

 In principle models can inform  $s$-dependence in background scattering (see, for example Ref.~\cite{Hilbert:2007hc}); however, we choose to implement the simplest and most agnostic model as:

\be
 f_{\mu ,\nu} = \hat{f}_{\mu , \nu}\cdot 
 \left(\frac{k_\mu(s)}{\beta}\right)^{\ell_\mu} \cdot 
 \left(\frac{k_\nu(s)}{\beta}\right)^{\ell_\nu}\cdot \exp\left[-\left(\frac{ k^2_\nu(s)+ k^2_\mu(s)}{\beta^2}\right)\right].
\label{eq:fFormFactor}
\ee
This form was chosen for simplicity, consistency with the coupling model, and because we found that some sort of high energy damping was beneficial to fit robustness. The model was used for all two-body background channels except $e^+e^-$, where a form factor is inappropriate.

Finally, we report useful formulae for extracting physical quantities. All of these results follow the conventions of the RPP with the sole exception of the factor of  $(2\pi)^4$ in the invariant phase space. Thus the differential cross section and decay width are given by

\be
d\sigma = \frac{|\mathcal{M}|^2}{4 p_i \sqrt{s}} d\Phi_n
\ee
and
\be
d\Gamma = \frac{|\mathcal{M}|^2}{2M} d\Phi_n.
\ee
The cross section for exclusive two-body $e^+e^-$ annihilation to channel $\mu$ is given by
\be
\sigma(e^+e^- \to \mu) =  \frac{1}{16 \pi s} \frac{k_\mu}{k_{ee}} \overline{|\M_{\mu,ee}|^2}.  
\label{eq:HMSXsec}
\ee
For two-to-three scattering, $12 \to 345$,  the differential cross section is given by: 
\be
d \sigma = \frac{1}{(2\pi)^3}\frac{1}{64 s \sqrt{s} k_i}\overline{|\M_{345:12}|^2} \, dm_{12}^2 \, dm_{23}^2. \label{eq:HMSXsec2}
\ee
Finally, if the vector decay constant is defined by

\be
\langle 0|\bar \psi \gamma^\mu \psi|V\rangle = m_V f_V \epsilon^\mu
\ee
then the perturbative relationship to the coupling is given by
\be
f_V = \sqrt{3} \, \left|\frac{g_{V:ee}(m_V^2)}{8\pi \alpha Q}\right|.
\ee

Similarly, the perturbative three-body partial width is given by

\be
\Gamma_{R:\Delta} = \frac{g_{R:\Delta}^2 A_{R:\Delta}(m_R^2)}{32(2\pi)^3 m_R^3}
\ee
where $A$ is the area of the Dalitz plot for this decay.

In this work we do not rely on perturbative expressions, rather partial widths are extracted from residues, as we have described in Sec.~\ref{sect:fit}. In this regard, two useful relationships are 
\be
\Gamma_{R:\Delta} = \frac{|Res(\mathcal{M}_{\mu,\Delta})|^2}{|Res(\mathcal{M}_{\mu,\mu})|} \cdot \frac{A_{R:\Delta}}{32 (2\pi)^3 m_R^3}
\ee
and, for the two-body case, 
\be
\Gamma_{R:\nu} = \frac{|Res(\mathcal{M}_{\mu,\nu})|^2}{|Res(\mathcal{M}_{\mu,\mu})|} \cdot \frac{\rho_\nu}{m_R}.
\ee

\section{Open Bottom Channels}
\label{appB}

We present fit results for the case in which only two-body channels are considered. This is of interest because the evidence for the $\Upsilon(10750)$ was in three-body decay modes~\cite{MIZUK19} and we wish to examine the possibility that this state can be discerned in the two-body data. 
We have therefore examined the two-body system with three models defined by $\beta = 0.8$, 1.0, and 1.2 GeV. Fit results are shown in Fig. \ref{fig:onlyBBfits}. We note that the general features of the fits are similar to those of the full model. Paucity of data, however, is evidenced by nearly degenerate minima that are found for the $\beta=  0.8$ and 1.0 GeV cases. Global $\chi^2$ values are given in Tab. \ref{tab:chi2b}, where we see very similar results (if slightly worse) to the full case.

\begin{table}[ht]
\begin{tabular}{l|ccc}
\hline\hline
model & $\chi^2$ & ndf & $\chi^2$/ndf \\
\hline
$\beta=0.8$/ 2-body/ (1) & 301 & 257 & 1.17 \\
$\beta=0.8$/ 2-body/ (2) & 301 & 257 & 1.17 \\
$\beta=1.0$/ 2-body/ (1) & 309 & 257 &  1.20 \\
$\beta=1.0$/ 2-body/ (2) & 307 & 257 & 1.19 \\
$\beta=1.2$/ 2-body & 298 & 257 & 1.16 \\
\hline\hline
\end{tabular}
\caption{Global $\chi^2$ values for the different open bottom-only fit models. In case we find more than one solution with a similar global $\chi^2$, both solutions are given.}
\label{tab:chi2b}
\end{table}

Pole positions for the restricted model are given in Fig. \ref{fig:onlyBBpoles}. Once again, results are very close to those of the full model.

These results yield strong evidence for the $\Upsilon(10750)$, with a similar range of resonance parameters as when three-body channels are included in the dataset. 
The minimum $\chi^2$ in fits with three poles
is at least 150 units worse with 6 fewer parameters than in fits with four poles,
again leading to a statistical significance of larger than 10~$\sigma$.
We conclude that the evidence for the $\Upsilon(10750)$ is robust and convincing.

\newpage
\begin{widetext}
$\quad$
\begin{figure}[h!]
    \centering
    \begin{overpic}[width=1.0\textwidth]{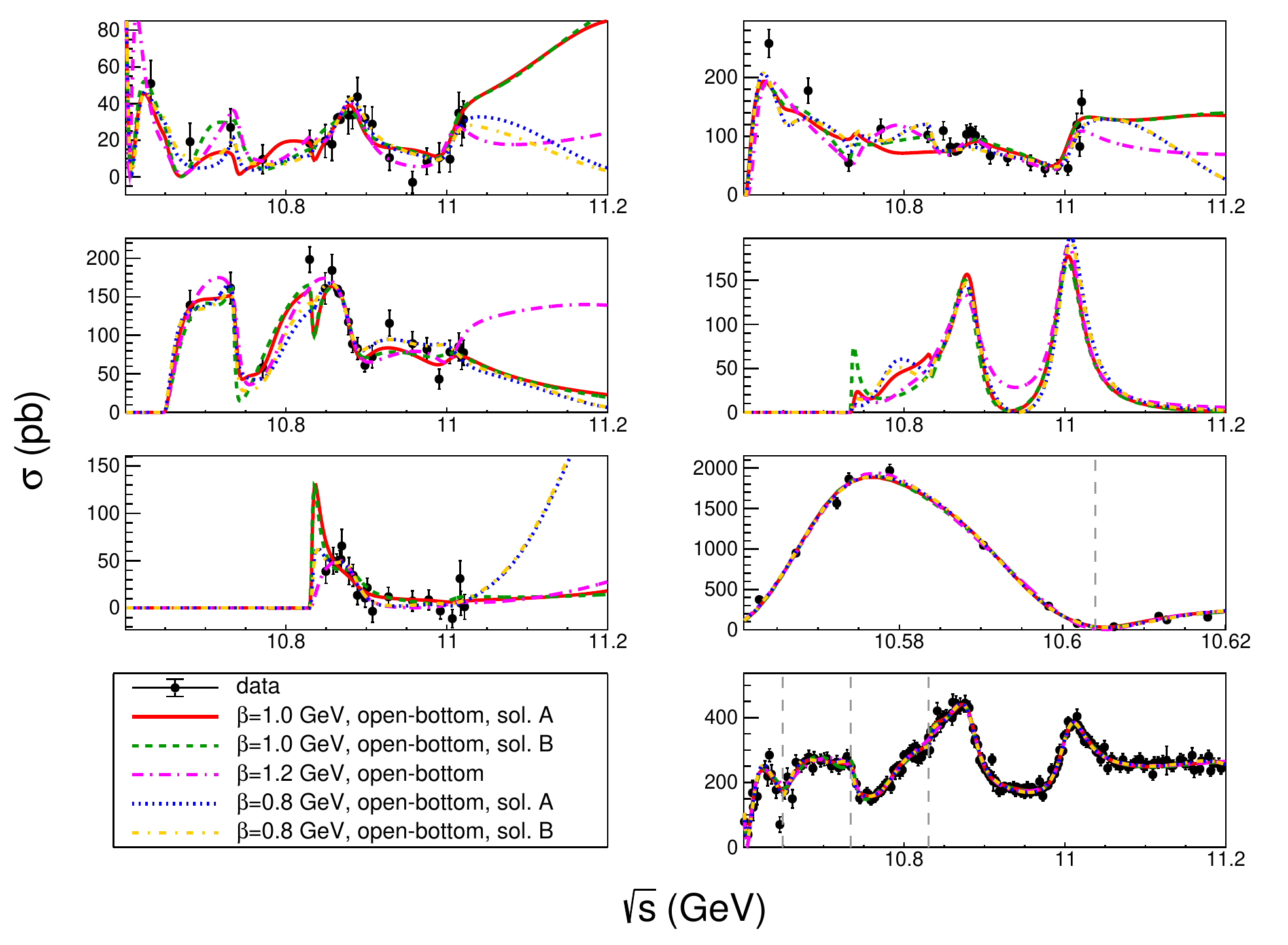}
    \put(32,69.5){(a) $B\bar{B}$}
    \put(86,69.5){(b) $B^*\bar{B}$}
    \put(32,52.5){(c) $B^*\bar{B}^*$}
    \put(85,52.5){(d) ``$B_s\bar{B}_s$"}
    \put(32,35.5){(e) $B_s^*\bar{B}_s^*$}
    \put(86,35.5){(f) $\sigma_{b\bar{b}}$}
    \put(86,18.5){(g) $\sigma_{b\bar{b}}$}
    \end{overpic}
    \caption{Fit results using only the two-body data for three different values of $\beta$. Ambiguous solutions for two of the three models are also shown.}
    \label{fig:onlyBBfits}
\end{figure}

\begin{figure}[h!]
    \centering
    \includegraphics[width=0.7\textwidth]{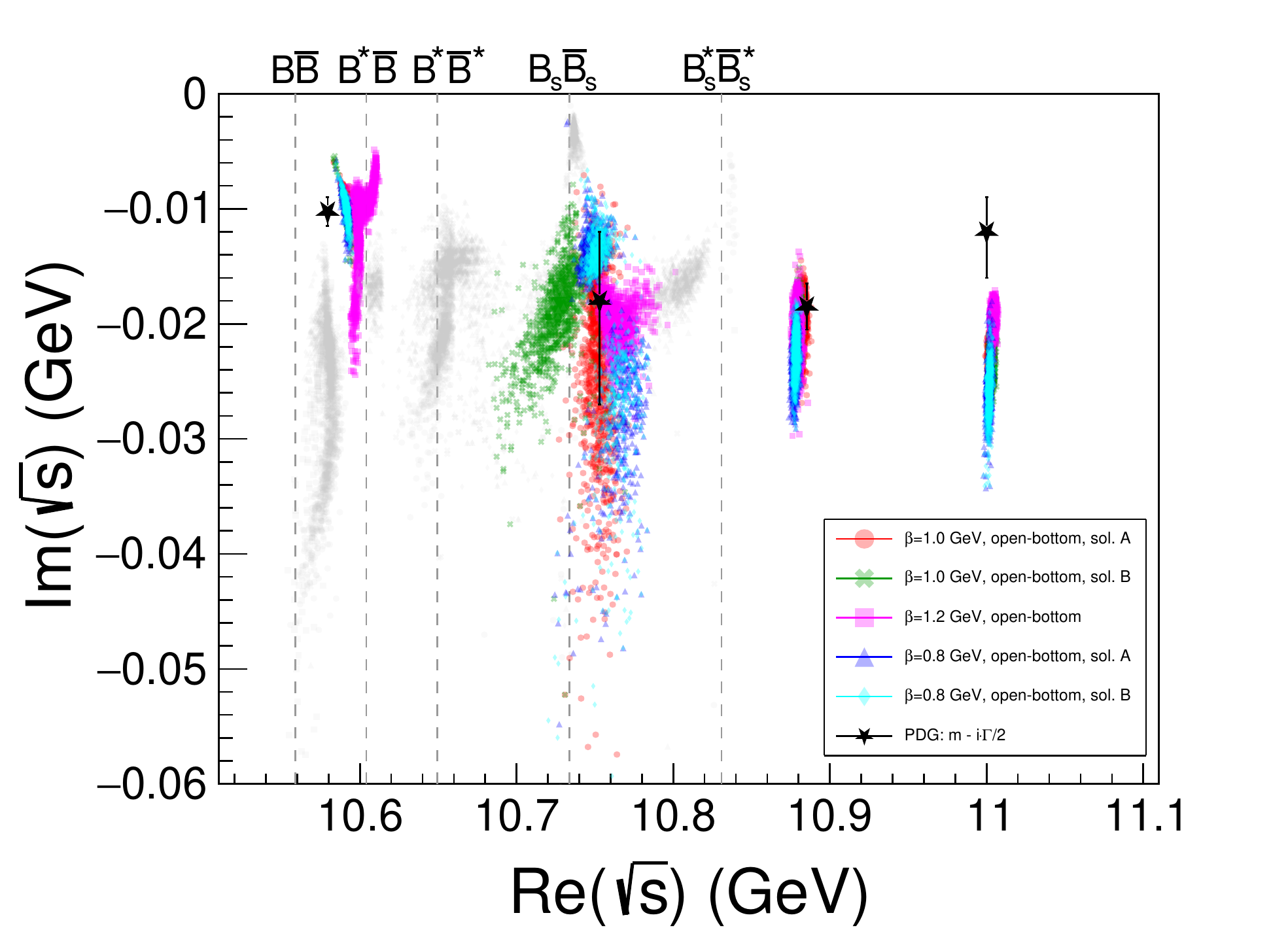}
    \caption{Bootstrap pole positions for three different models and two further ambiguous solutions indicated by different colors and markers. Gray markers indicate ghost poles with sizable residues in that model, black stars indicate the RPP average values using a Breit-Wigner parameterization.}
    \label{fig:onlyBBpoles}
\end{figure}
\end{widetext}


\begin{thebibliography}{0}

\bibitem{BESSON85}
D.~Besson \textit{et al.} [CLEO],
Phys. Rev. Lett. \textbf{54}, 381 (1985).

\bibitem{LOVELOCK85}
D.~M.~J.~Lovelock, J.~E.~Horstkotte, C.~Klopfenstein, J.~Lee-Franzini, L.~Romero, R.~D.~Schamberger, S.~Youssef, P.~Franzini, D.~Son and P.~M.~Tuts, \textit{et al.}
Phys. Rev. Lett. \textbf{54}, 377-380 (1985).

\bibitem{AUBERT09E}
B.~Aubert \textit{et al.} [BaBar],
Phys. Rev. Lett. \textbf{102}, 012001 (2009)
[arXiv:0809.4120 [hep-ex]].

\bibitem{SANTEL16}
D.~Santel \textit{et al.} [Belle],
Phys. Rev. D \textbf{93}, no.1, 011101 (2016)
[arXiv:1501.01137 [hep-ex]].

\bibitem{DONG20A}
X.~K.~Dong, X.~H.~Mo, P.~Wang and C.~Z.~Yuan,
Chin. Phys. C \textbf{44}, no.8, 083001 (2020)
[arXiv:2002.09838 [hep-ph]].

\bibitem{MIZUK21bb}
R.~Mizuk \textit{et al.} [Belle],
JHEP \textbf{06}, 137 (2021)
[arXiv:2104.08371 [hep-ex]].


\bibitem{pdg}
P.A. Zyla \textit{et al.} (Particle Data Group), Prog. Theor. Exp. Phys. 2020, 083C01 (2020) and 2021 update. 




\bibitem{AQUINES06}
O.~Aquines \textit{et al.} [CLEO],
Phys. Rev. Lett. \textbf{96}, 152001 (2006)
[arXiv:hep-ex/0601044 [hep-ex]].

\bibitem{HUANG07}
G.~S.~Huang \textit{et al.} [CLEO],
Phys. Rev. D \textbf{75}, 012002 (2007)
[arXiv:hep-ex/0610035 [hep-ex]].

\bibitem{DRUTSKOY10}
A.~Drutskoy \textit{et al.} [Belle],
Phys. Rev. D \textbf{81}, 112003 (2010)
[arXiv:1003.5885 [hep-ex]].

\bibitem{CHEN08}
K.~F.~Chen \textit{et al.} [Belle],
Phys. Rev. Lett. \textbf{100}, 112001 (2008)
[arXiv:0710.2577 [hep-ex]].

\bibitem{CHEN10}
K.~F.~Chen \textit{et al.} [Belle],
Phys. Rev. D \textbf{82}, 091106 (2010)
[arXiv:0810.3829 [hep-ex]].

\bibitem{ADACHI12}
I.~Adachi \textit{et al.} [Belle],
Phys. Rev. Lett. \textbf{108}, 032001 (2012)
[arXiv:1103.3419 [hep-ex]].

\bibitem{MIZUK16}
A.~Abdesselam \textit{et al.} [Belle],
Phys. Rev. Lett. \textbf{117}, no.14, 142001 (2016)
[arXiv:1508.06562 [hep-ex]].

\bibitem{MIZUK19}
R.~Mizuk \textit{et al.} [Belle],
JHEP \textbf{10}, 220 (2019)
[arXiv:1905.05521 [hep-ex]].

\bibitem{BONDAR12}
A.~Bondar \textit{et al.} [Belle],
Phys. Rev. Lett. \textbf{108}, 122001 (2012)
[arXiv:1110.2251 [hep-ex]].

\bibitem{GARMASH15}
A.~Garmash \textit{et al.} [Belle],
Phys. Rev. D \textbf{91}, no.7, 072003 (2015)
[arXiv:1403.0992 [hep-ex]].

\bibitem{albrecht}
M. Albrecht \textit{et al.}, Eur. Phys. J. C \textbf{80} (2020) 5, 453

\bibitem{JPAC}
A.~Rodas \textit{et al.} [Joint Physics Analysis Center],
Eur. Phys. J. C \textbf{82} (2022) no.1, 80
doi:10.1140/epjc/s10052-022-10014-8
[arXiv:2110.00027 [hep-ph]].

\bibitem{BONN}
A.~V.~Sarantsev, I.~Denisenko, U.~Thoma and E.~Klempt,
Phys. Lett. B \textbf{816} (2021), 136227
doi:10.1016/j.physletb.2021.136227
[arXiv:2103.09680 [hep-ph]].


\bibitem{Uglov:2016orr}
T.~V.~Uglov, Y.~S.~Kalashnikova, A.~V.~Nefediev, G.~V.~Pakhlova and P.~N.~Pakhlov,
JETP Lett. \textbf{105}, no.1, 1-7 (2017)
[arXiv:1611.07582 [hep-ph]].

\bibitem{ian}
I.J.R. Aitchison, Nucl. Phys. {\bf A189}, 417 (1972).

\bibitem{3bodyK}
G. Ascoli and H. W. Wyld,
Phys. Rev. D 12, 43 (1975); 
M. Mai, B. Hu, M. D\"{o}ring, A. Pilloni and A. Szczepaniak, EPJA 53, 177 (2017);
A.W. Jackura, S.M. Dawid, C. Fern\'{a}ndez-Ram\'{i}rez, V. Mathieu, M. Mikhasenko, A. Pilloni, S.R. Sharpe, and A.P. Szczepaniak,
Phys. Rev. D 100, 034508(2019).

\bibitem{Belle:2006jvm}
A.~Drutskoy \textit{et al.} [Belle],
Phys. Rev. Lett. \textbf{98} (2007), 052001
doi:10.1103/PhysRevLett.98.052001
[arXiv:hep-ex/0608015 [hep-ex]].



\bibitem{BelleBsBs}
A.~Abdesselam \textit{et al.} [Belle],
arXiv:1609.08749


\bibitem{Belleon4S}
E. Guido \textit{et al.} [Belle],
Phys. Rev. D \textbf{96}, 052005 (2017).

\bibitem{MizukPriv}
R.~Mizuk, private communication

\bibitem{mizukTalk}
R. Mizuk, contribution to QWG2021 - The 14th International Workshop on Heavy Quarkonium,  \url{https://indico.cern.ch/event/838970/contributions/4260186/attachments/2210504/3740978/mizuk_QWG2021.pdf}


\bibitem{Hilbert:2007hc}
J.~P.~Hilbert, N.~Black, T.~Barnes and E.~S.~Swanson,
Phys. Rev. C \textbf{75}, 064907 (2007)
doi:10.1103/PhysRevC.75.064907
[arXiv:nucl-th/0701087 [nucl-th]].

\bibitem{Godfrey:2015dia}
S.~Godfrey and K.~Moats,
Phys. Rev. D \textbf{92}, no.5, 054034 (2015)
doi:10.1103/PhysRevD.92.054034
[arXiv:1507.00024 [hep-ph]].


\bibitem{Segovia:2016xqb}
J.~Segovia, P.~G.~Ortega, D.~R.~Entem and F.~Fern\'andez,
Phys. Rev. D \textbf{93}, no.7, 074027 (2016)
doi:10.1103/PhysRevD.93.074027

\bibitem{Barnes:2005pb}
T.~Barnes, S.~Godfrey and E.~S.~Swanson,
Phys. Rev. D \textbf{72}, 054026 (2005)
doi:10.1103/PhysRevD.72.054026
[arXiv:hep-ph/0505002 [hep-ph]].

\bibitem{Godfrey:2016nwn}
S.~Godfrey, K.~Moats and E.~S.~Swanson,
Phys. Rev. D \textbf{94}, no.5, 054025 (2016)
doi:10.1103/PhysRevD.94.054025
[arXiv:1607.02169 [hep-ph]].

\bibitem{Farina:2020slb}
C.~Farina, H.~Garcia Tecocoatzi, A.~Giachino, E.~Santopinto and E.~S.~Swanson,
Phys. Rev. D \textbf{102}, no.1, 014023 (2020)
doi:10.1103/PhysRevD.102.014023
[arXiv:2005.10850 [hep-ph]].

\bibitem{Ryan:2020iog}
S.~M.~Ryan \textit{et al.} [Hadron Spectrum],
JHEP \textbf{02}, 214 (2021)
doi:10.1007/JHEP02(2021)214
[arXiv:2008.02656 [hep-lat]].

\bibitem{Lakhina:2006vg}
O.~Lakhina and E.~S.~Swanson,
Phys. Rev. D \textbf{74}, 014012 (2006)
doi:10.1103/PhysRevD.74.014012
[arXiv:hep-ph/0603164 [hep-ph]].

\bibitem{Chen:2019gty}
Y.~H.~Chen and F.~K.~Guo,
Phys. Rev. D \textbf{100}, no.5, 054035 (2019)
doi:10.1103/PhysRevD.100.054035
[arXiv:1906.05766 [hep-ph]].

\bibitem{TarrusCastella:2021pld}
J.~Tarr\'us Castell\`a and E.~Passemar,
Phys. Rev. D \textbf{104}, no.3, 034019 (2021)
doi:10.1103/PhysRevD.104.034019
[arXiv:2104.03975 [hep-ph]].


\bibitem{Wilson:2014cna}
M.R. Pennington personal communication; as cited in 
D.~J.~Wilson, J.~J.~Dudek, R.~G.~Edwards and C.~E.~Thomas,
Phys. Rev. D \textbf{91}, no.5, 054008 (2015)
doi:10.1103/PhysRevD.91.054008
[arXiv:1411.2004 [hep-ph]].
See also
J.L. Basdevant and E.L. Berger, Phys. Rev. D \textbf{16}, 657 (1977).

\bibitem{rpp50}
See, for example, Section 50 of~\cite{pdg}.

\bibitem{Gribov:2009zz}
V.~N.~Gribov, Y.~L.~Dokshitzer and J.~Nyiri,
``Strong interactions of hadrons at high energies: Gribov lectures on 
Theoretical Physics''.

\bibitem{B}
G. Breit, Phys. Rev. {\bf 69}, 472 (1946);
H. Feshbach, D.C. Peaslee, and V. Weisskopf, Phys. Rev. {\bf 71}, 145 (1947); M. Gell-Mann and K.M. Watson, Ann. Rev. Nuc. Sci. {\bf 4}, 219 (1954); H. Pilkuhn, {\sl Relativistic Particle Physics}, pg 171 (Springer-Verlag, New York, 1979), pg. 171.


\bibitem{Walker} R.L. Walker, Phys. Rev. {\bf 182}, 1729 (1969).

\bibitem{maid} D. Drechsel, O. Hanstein, S.S. Kamalov,
and L. Tiator, Nucl. Phys. A{\bf 645}, 145 (1999).

\bibitem{vHQ}
F.~Von Hippel and C.~Quigg,
Phys.\ Rev.\ D {\bf 5}, 624 (1972).

\bibitem{BlW}
J. Blatt and V. Weisskopf, {\sl Theoretical Nuclear Physics}, pg. 358 (Dover,
New York, 1991); \bibitem{VonHippel:fg}
F.~Von Hippel and C.~Quigg,
Phys.\ Rev.\ D {\bf 5}, 624 (1972).


\bibitem{BaBar:2004rrm}
B.~Aubert \textit{et al.} [BaBar],
Phys. Rev. D \textbf{72}, 032005 (2005)
doi:10.1103/PhysRevD.72.032005
[arXiv:hep-ex/0405025 [hep-ex]].



















\end{thebibliography}
\end{document}